\documentclass[12pt]{iopart}
\usepackage{iopams}

\usepackage[T1]{fontenc}
\usepackage{pxfonts}

\usepackage{epsfig,color}
\usepackage{epstopdf}

\begin{document}

\def\[{\begin{equation}}
\def\]{\end{equation}}
\def\_#1{{\bf #1}}
\def\o{\omega}
\def\.{\cdot}
\def\x{\times}
\def\E{\varepsilon}
\def\M{\mu}
\newcommand{\ds}{\displaystyle}
\def\l#1{\label{eq:#1}}
\def\r#1{(\ref{eq:#1})}
\def\no{\noindent}
\def\d{\partial}

\def\ds{\displaystyle}
\def\ss{\scriptstyle}

\renewcommand\Re{{\rm Re}}
\renewcommand\Im{{\rm Im}}

\newcommand\vect[1]{\left(\begin{array}{r}#1\end{array}\right)}
\newcommand\matr[1]{\left(\begin{array}{cc}#1\end{array}\right)}

\title[Overcoming black body radiation limit in free space]{Overcoming black body radiation limit in free space: metamaterial superemitter}

\author{Stanislav I. Maslovski$^1$, Constantin R. Simovski$^2$, Sergei A. Tretyakov$^2$}

\address{$^1$ Departamento de Engenharia Electrot\'{e}cnica\\
Instituto de Telecomunica\c{c}\~{o}es, Universidade de Coimbra\\
P\'{o}lo II, 3030-290 Coimbra, Portugal
}
\address{$^2$ Aalto University, School of Electrical Engineering\\
P.O. Box 13000, 00076 Aalto, Finland
}
\ead{stas@co.it.pt}

\begin{abstract}
  Here, we demonstrate that the power spectral density of thermal
  radiation at a specific wavelength produced by a body of finite
  dimensions set up in free space under a fixed temperature could be
  made theoretically arbitrary high, if one could realize double
  negative metamaterials with arbitrary small loss and arbitrary high
  absolute values of permittivity and permeability (at a given
  frequency). This result refutes the widespread belief that Planck's
  law itself sets a hard upper limit on the spectral density of power
  emitted by a finite macroscopic body whose size is much greater that
  the wavelength. Here we propose a physical realization of a
  metamaterial emitter whose spectral emissivity can be greater than
  that of the ideal black body under the same conditions. Due to the
  reciprocity between the heat emission and absorption processes such
  cooled down superemitter also acts as an optimal sink for the
  thermal radiation --- the ``thermal black hole'' --- which
  outperforms Kirchhoff-Planck's black body which can absorb only the
  rays directly incident on its surface. The results may open a
  possibility to realize narrowband super-Planckian thermal radiators
  and absorbers for future thermo-photovoltaic systems and other devices.\\[2mm]
  \noindent{\it Keywords\/}: metamaterial, thermal radiation, black body.
\end{abstract}

\pacs{44.40.+a,  78.67.Pt, 42.25.Fx}

\maketitle

\section{Introduction}

The ability of a hot body to emit thermal electromagnetic radiation is
related to its ability to absorb incident electromagnetic waves at the
same frequencies.  G.~Kirchhoff in 1860 introduced the theoretical
concept of an ideal black body, which ``completely absorbs all
incident rays''~\cite{GK}. This concept was later adopted by
M.~Planck~\cite{Planck}. It appears that since that time there has been a
general belief that no macroscopic body can emit more thermal
radiation than the corresponding same-shape and size ideal black body
at the same temperature.  For example, in a recent paper~\cite{4} one
reads, ``any actual macroscopic thermal body cannot emit more thermal
radiation than a blackbody.''  Equivalent statements can be found in
commonly used text books, for example, in the well-known book by
Bohren and Huffman~\cite{Bohren} it is stated that ``\dots the
emissivity of a sufficiently large sphere is not greater than 1. Thus,
if the radiating sphere radius is much larger than the wavelength, the
radiation above the black body limit is impossible.''

On the other hand, recently there has been increasing number of
publications discussing so-called super-Planckian thermal radiation,
when the power emitted by a hot body per unit area per unit wavelength
exceeds the one predicted by Planck's black body law.  In a great deal
of such works, the thermal emission into the electromagnetic {\em
  near-field} is considered, when the bodies that exchange radiative
heat are separated by a distance significantly smaller than the
wavelength $\lambda$ (on the order of $\lambda/10$ or less). Such
emission can easily overcome the black body limit, because oscillators
in bodies separated by subwavelength gaps interact through the near
(i.e., Coulomb) electric field, and, when close enough to the emitting
object, such a field is much stronger than the wave field.

Besides the near field transfer, there are also works --- quite
surprising for an unprepared reader --- which report super-Planckian
emission in {\it far-field in free-space}.  In order to avoid
confusion we must first agree on the terminology, because it appears
that, currently, in the literature there is no consensus on the meaning of
the super-Planckian radiation in this case.  In this paper, we use
this terminology {\em exclusively for bodies of finite dimensions}.
When such an object acts as a source of thermal radiation, its
spectral radiance $b_{\lambda}$, i.e., the amount of power $d^2P$
radiated per wavelength interval $d\lambda$, projected emitting area
$A_{\perp}$, and solid angle $d\Omega$:
\[
b_{\lambda} = {d^2P\over A_{\perp}d\lambda\,d\Omega},
\l{radiance}
\]
is sub-Planckian or super-Planckian {\em depending on the choice} of
the area $A_{\perp}$. If {\em by definition} $A_{\perp}$ is chosen to
coincide with emitter's geometric projected area: $A_{\perp} = A_{\rm
  geom}$, then, given the reciprocal nature of the radiative heat
emission and absorption processes, one has to admit that the spectral
radiance of bodies characterized with the effective absorption cross
section $\sigma_{\rm abs}$ such that $\sigma_{\rm abs} > A_{\rm
  geom}$ must be {\rm super-Planckian}, because for the ideal black
body $\sigma_{\rm abs} = A_{\rm geom}$.

Such definition explains the known fact that an {\em optically small}
body can emit more than a black body of a similar size. Indeed, a
small particle may absorb much more power than one would expect from
its size, because a particle with radius $a \ll \lambda$ may have the
absorption cross section $\sigma_{\rm abs}$ much larger than its
geometric cross section $A_{\rm geom} = \pi a^2$. For a dielectric or
plasmonic sphere, this is understood as a consequence of Mie's (or
respective plasmonic) resonances, at which the absorption cross
section $\sigma_{\rm abs}$ may outnumber $\pi a^2$ by a large factor.
For instance, for a single-mode dipole particle the ultimate
absorption cross section equals $\sigma_{\rm abs\, max}=({3/
  8\pi})\lambda^2$ (e.g.~Ref.~\cite{maximizing}), which is much larger
than $\pi a^2$ if $a\ll \lambda$. The absorption cross section can be
further increased if the incident field couples to many resonant
modes,
e.g. Refs.~\cite{Bobr,unlimited_antenna,Fan,Alu2014}. Similarly,
resonant absorption by shape irregularities with curvature radius
$a\ll \lambda$ on a surface of a large body~\cite{geometric} makes the
absorption cross section associated with an irregularity larger than
its geometric cross section.

However, the known literature does not provide a definite
answer to the main question of this article, namely, up to which
degree the spectral density of power emitted by an {\em optically
  large body} can be larger than the one produced by the black body of
the same dimensions under the same thermal conditions? In other words,
how prominent can be the super-Planckian free-space radiation effect
mentioned above in bodies whose size is {\em much greater than the
  wavelength?}

In fact, there is no agreement in the current literature even on
whether such super-Planckian emission into free space is physically
allowed --- in particular, by thermodynamical considerations, --- in a
scenario with an optically large body emitting. For instance, in
Ref.~\cite{2nd_law} it is argued that such emission would violate the
second law of thermodynamics, but is this indeed the case?

In this paper, unlike previous works on related
subjects, we consider a theoretical possibility to obtain
\emph{free-space omnidirectional super-Planckian radiation from a
  finite macroscopic body with characteristic radius $a \gg
  \lambda$}. This implies an important question --- if there can exist
optically large isotropic emitters with effective spectral emissivity
greater than unity, when compared to Kirchhoff-Planck's black body of
the same size.  In order to answer this question, we go
over the usual assumption that an emitting body is composed of
homogeneous materials with {\em positive} permittivities and
permeabilities at the wavelength of interest. We show that if these
restrictions are removed, there are no compelling reasons why a
specially crafted {\em metamaterial}~\cite{Engheta} object cannot
produce a higher radiated spectral power at a given wavelength than
the respective black body, even when object's diameter is
significantly greater than the wavelength.

Moreover, here we prove that the spectral power produced by a
double-negative metamaterial emitter can be made theoretically {\em
  arbitrary} high at any given frequency, {\em independently} of the
physical size of the emitter, under the condition that arbitrarily low
loss tangent values and arbitrarily high absolute values of the
permittivity and permeability are attainable. When cooled, such objects
act as ``thermal black holes'' which absorb much more power than is
incident directly on their surfaces. For instance, in a plane wave
illumination scenario, they absorb (theoretically) the whole infinite
power carried by such a wave (of infinite extent in space).

We prove that existence of such superemitters (and superabsorbers)
contradicts neither the second law of thermodynamics, nor Kirchhoff's
law of thermal radiation when the latter is properly amended.
In particular, we show that the super-Planckian part of the thermal
flux in the vicinity of a superemitter is transferred by resonant
tunneling of photons associated with high-order, highly reactive
spatial harmonics (essentially, dark modes) of emitter's fluctuating
field.
Due to this process, the effective spectral emissivity of such emitter
when compared to Kirchhoff-Planck's black body of the same size is
{\em greater than unity}. This is physically possible because
effective absorptivity of this object is {\em as well greater than
  unity,} which just means that it receives per unit area of its
surface more spectral power than a Kirchhoff-Planck's black body of
the same radius under the same conditions.

Note that earlier studies establishing the widely accepted limitations
are based on an assumption that the ideal Kirchhoff-Planck black body
is the ultimately effective absorber. However, such a body perfectly
absorbs only the rays which are falling directly on its
surface~\cite{GK}. Furthermore, we may recall the known result from
the electromagnetic theory which states that there is no upper limit
on the effective area of an antenna, even when the physical dimensions
of the antenna are constrained~\cite{unlimited_antenna,Sch}, and the
equivalent results in diffraction theory \cite{Leman} and acoustics
\cite{Bobr}. In particular, this means that a finite-size antenna
loaded with a conjugate-matched load $Z_{\rm load} = Z_{\rm ant}^*$,
where $Z_{\rm ant}$ is the complex input impedance of the antenna, in
principle, can absorb all the power carried by a plane wave incident
from the direction of antenna's main beam, and thus --- for this
direction of incidence --- can be infinitely more efficient in
absorption than the ideal black body.

Similarly, the resonant photon tunneling effect which enables
super-Planckian radiation occurs when the emitter is
conjugate-impedance matched to a large set of the free space modes,
including both bright (propagating) and dark (nearly evanescent)
modes. As will be seen from the following, this effect is inherently
narrowband due to highly reactive nature of the electromagnetic field
associated with the dark modes.
Let us note, however, that narrowband thermal radiation is the key
prerequisite for advanced thermo-photovoltaic systems (TPVS). For
instance, reducing relative bandwidth to less than 10\% practically
eliminates the Shottky-Queisser limit related to the dissipation of
the excessive photon energy in semiconductors \cite{Wernsman}. The
nearly monochromatic thermal radiation is the primary target for solar
TPVS, where the narrowband thermal emitters already allow the energy
conversion efficiency to approach the thermodynamic limit
\cite{Chester}.  Note that in all known works on solar TPVS the
spectral maximum of this narrow-band thermal radiation is below the
conventional Planckian spectral value~\cite{Zoysa}.

In contrast, when a wide-band gain in thermal emission is needed, it
can be achieved by covering emitters with transparent dielectric
shells~\cite{4}, or with shells made of hyperbolic
metamaterials~\cite{arxivHyperlens}. The radiation enhancement in
these cases can be explained by an increase in the Purcell factor
associated with elementary sources of thermal radiation placed inside
such shells~\cite{arxivHyperlens}.
However, note that such non-resonant, broadband enhancers which
essentially operate as optical collimators may not increase the
effective emitter size beyond the size of the transparent shell
itself. This means that the emitted power in such systems never
exceeds the power radiated by a black body with the radius equal to
the outer radius of the shell. Thus, the thermal radiation flux
produced at the output of the shell is sub-Planckian.
Furthermore, thermal radiation from an unbounded planar interface with
a generic photonic crystal has been numerically studied in
Ref.~\cite{Luo}, and the results show that the power radiated from an
infinite planar surface does not go over the black body limit.

Thus, it is important to investigate, both from the theoretical and
practical points of view, the possibilities in realizing
omnidirectional thermal super-Planckian emitters and confirm that
their existence does not violate fundamental laws of physics, as well
as to find the required properties of metamaterials from which such
thermal superemitters can be made. These are the goals of this work.

\section{The content of this paper}

The paper is organized as follows. In Sec.~\ref{sectmodel} we outline
the equivalent circuit model~\cite{Maslovski_circuit_PRB_2013} that we
use in radiative thermal flux calculations. It has been
proven~\cite{Maslovski_circuit_PRB_2013} that this approach is fully
equivalent in its predictive power to the more common theories
operating with distributed thermal-fluctuating currents. Using this
model, in Sec.~\ref{sectconj} we consider general conditions which
maximize the radiative heat flux between a hot body and its
environment.

In Sec.~\ref{cmemitter} we study the thermal radiation produced by
finite-size bodies in free space and introduce the concept of the
ideal conjugate-matched emitter. Such a truly super-Planckian emitter
is able to radiate efficiently to the {\em entire} infinite set of
free-space photonic states, infinitely outperforming a black
body emitter of the same dimensions.

In Sec.~\ref{scattering} we consider plane wave scattering on a
finite-size body and prove that its scattering, absorption, and
extinction cross sections tend to infinity under the perfect conjugate
matching condition of Sec.~\ref{cmemitter}, independently of the size
of the body.

In Sec.~\ref{kirchhoff} we show that the second law of thermodynamics
is not violated by finite-size emitters with effective spectral
emissivity greater than unity. We also propose an amendment to
Kirchhoff's law of thermal radiation in order to incorporate such
emitters into the existing theory.

In Sec.~\ref{realization} we search for a physical realization for the
conjugate-matched emitter. A possible realization --- which we call
metamaterial thermal black hole --- is obtained in the form of a
core-shell double-negative (DNG) metamaterial structure.

In Sec.~\ref{results} we consider a couple of such structures with
realistic material parameters and estimate their super-Planckian
performance. Finally, in Sec.~\ref{conclusion} we draw some
conclusions.

\section{\label{sectmodel}Electromagnetic theory of thermal radiation: circuit model approach}

\begin{figure}[tb]
\centering
\epsfig{file=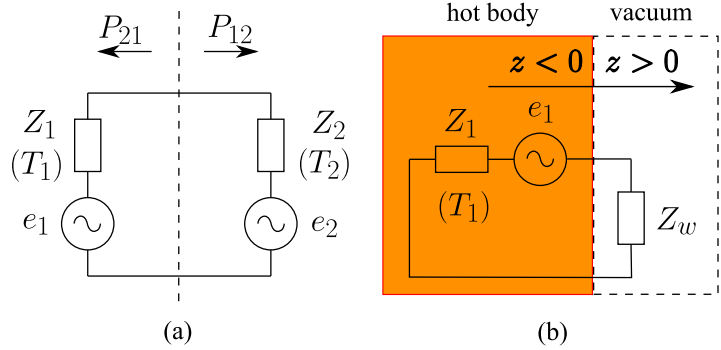,width=0.6\textwidth}
\caption{\label{circuit}
  (a) Equivalent circuit of radiative heat
  transfer between an emitter (represented by the complex
  impedance $Z_1$) and its environment (represented by the
  complex impedance $Z_2$). (b) Equivalent circuit for the particular case of
  an infinitely large hot body occupying the halfspace $z < 0$ and
  radiating into the cold free space domain $z > 0$. In this geometry, the
  impedance $Z_2 = Z_w$ is either purely real (for propagating plane waves)
  or purely imaginary (for evanescent waves).}
\end{figure}

The approach of Ref.~\cite{Maslovski_circuit_PRB_2013} allows
one to reduce a full-wave thermal emission problem to a set of circuit
theory problems operating with effective fluctuating voltages and
currents instead of the electromagnetic fields. This  approach
is based on expanding the emitted field at a given frequency into a
suitable series of linearly independent, orthogonal spatial harmonics,
and characterizing each of these harmonics with the equivalent circuit
model parameters, such as complex wave impedance, voltage and
current. The electromagnetic interaction of a hot emitter with the
surrounding space can be expressed in this language at each of the
mentioned harmonics with an equivalent circuit shown in
Fig.~\ref{circuit}(a). In this circuit, $Z_1(\nu)$ represents the
equivalent complex impedance of emitter's body for a given spatial
harmonic of the radiated field, at the frequency~$\nu =
c/\lambda$. Respectively, $Z_2(\nu)$ is the equivalent complex
impedance of the surrounding space for the same mode, which, in case
of the free space, is simply the wave impedance of the corresponding
mode: $Z_2 \equiv Z_w$. The effect of thermal fluctuations in this
circuit is taken into account by a pair of fluctuating electromotive
forces (EMF) $e_1(\nu)$ and $e_2(\nu)$.

For example, in a geometry where a body occupying halfspace $z < 0$
emits into empty halfspace $z > 0$ [Fig.~\ref{circuit}(b)] one may
conveniently expand the radiated field over the set of free-space
plane waves (both propagating and evanescent) with arbitrary
transverse wave vectors $\_k_{\rm t} = (k_x,k_y)$. Such modes split
into transverse electric (TE) (or $s$-polarized) waves and transverse
magnetic (TM) (or $p$-polarized) waves. The wave impedances of these
modes $Z_w^{\rm TE,TM}(\nu, q)$ (where $q = |\_k_{\rm t}|$) satisfy
\[
Z_w^{\rm TE} =
{\eta_0\over\sqrt{1-q^2/k_0^2}}, \quad Z_w^{\rm TM} = \eta_0\sqrt{1-q^2/k_0^2},
\l{zwave}
\]
where $k_0=2\pi\nu\sqrt{\E_0\M_0}$ is the free-space wavenumber, and
$\eta_0=\sqrt{\M_0/\E_0}$ is the free-space impedance. Respectively,
in the equivalent circuit of Fig.~\ref{circuit}(b), $Z_2 = Z_w^{\rm
  TE,TM}$.

The impedance $Z_1$ in this case coincides with the input impedance of
the halfspace $z < 0$ for a given plane wave incident from the
halfspace $z > 0$. This impedance can be expressed through the
corresponding complex reflection coefficients $\Gamma_{\rm TE,TM}$ as
\[
Z_1 = Z_w^{\rm TE,TM}{1+\Gamma_{\rm TE,TM}\over 1-\Gamma_{\rm TE,TM}}.
\]

By applying the fluctuation-dissipation theorem~\cite{FDT} (FDT) to
the circuit of Fig.~\ref{circuit}(a) one finds the mean-square
spectral density of the fluctuating EMF as follows (in this article we
use rms complex amplitudes $x_\nu$ for the time-harmonic quantities
$x(t)$ defined by $x(t) = \Re[\sqrt{2}\,x_\nu \exp(-i\,2\pi\nu t)]$,
where $i = \sqrt{-1}$; therefore, $\overline{x^2} = |x_\nu|^2$):
\[
{d\overline{{e}^2_{j}}\over d\nu} = 4\Theta(\nu, T_j)\Re(Z_j),
\l{emf}
\]
where $\ds\Theta(\nu, T_j) = h\nu[\exp(h\nu/k_{\rm B}T_j)-1]^{-1}$ is
Planck's mean oscillator energy (here, $j = 1, 2$), $k_{\rm B}$ is Boltzmann's
constant, and $T_j$ is the absolute temperature of the emitter (when
$j = 1$) or the surrounding space (when $j = 2$). Actually,
Eq.~\r{emf} is nothing more than Nyquist's formula for the thermal
noise in electric circuits~\cite{Nyquist} where electrical
engineers usually approximate $\Theta(\nu,T_j)\approx k_{\rm B}T_j$.
Let us note that relation~\r{emf} implies that the bodies that
exchange radiative heat are kept in thermodynamically equilibrium
states, which, strictly speaking, is possible only either when $|T_1 -
T_2| \ll T_{1,2}$ or under the assumption that the internal thermal
energy stored in the bodies is infinite.

The thermal radiation power within a narrow range of frequencies
$\nu\pm d\nu/2$ delivered from the side of the emitter, $Z_1$, to the
side of the environment, $Z_2$, is expressed in our formulation (per
each spatial harmonic) simply as
\[
dP_{12} = {\Re(Z_2)\,d\overline{e^2_1}\over\left|Z_1+Z_2\right|^2}
= {4\Re(Z_1)\Re(Z_2)\over\left|Z_1+Z_2\right|^2}\,\Theta(\nu, T_1)\,d\nu.
\l{pow}
\]
In Ref.~\cite{Maslovski_circuit_PRB_2013} it is proven that such
a circuit model approach based on modal decomposition of the thermal
fluctuating field is fully equivalent to the more complicated theories
operating with distributed fluctuating currents. However, in contrast
to these classical methods, our approach allows us to reduce a heat
transfer maximization problem to the well-known circuit theory problem
of matching a generator with its load.

\section{\label{sectconj}Maximization of emitted power:
  complex-conjugate matching versus usual impedance matching}

Having at hand an equivalent circuit representation described above,
we may now ask ourselves under which conditions the spectral density
of power radiated by a hot body is maximized? Due to the orthogonality
property of the spatial harmonics used in the field expansion, in
order to maximize the total emission we need to maximize the power
delivered by each harmonic separately. As is clearly seen from
Eq.~\r{pow}, for the modes with a non-vanishing real part of the wave
impedance: $\Re(Z_2)>0$, the delivered power is maximized under the
complex-conjugate matching condition: $Z_1^*=Z_2$. Under this condition, the
maximal possible emitted power per a spatial harmonic per unit of
frequency is, from Eq.~\r{pow},
\[
{dP_{\rm max}\over d\nu} = \Theta(\nu,T_1).
\]

Note that for the spatial harmonics characterized with complex wave
impedance, the conjugate matching condition is, in general, different
from the zero reflection condition $\Gamma = {(Z_1-Z_2)/(Z_1+Z_2)} =
0$ in the equivalent circuit of Fig.~\ref{circuit}(a), which implies
the usual impedance matching $Z_1=Z_2$. Thus, by minimizing
reflections for the waves crossing the boundary between the emitter
and the surrounding space, one does not necessarily maximize the
emission! Indeed, the power spectral density under the usual impedance
matching condition $Z_1=Z_2$ (in what follows we call it simply
``impedance matching'') attains [Eq.~\r{pow}]
\[
{dP_{12}\over d\nu} = \left({\Re(Z_2)\over |Z_2|}\right)^2\Theta(\nu,T_1) \le
{dP_{\rm max}\over d\nu}.
\l{specdensbb}
\]
Recall that $Z_2$ is related to the wave impedance of the surrounding
space. Therefore, as soon as this environment is characterized with
{\em complex} impedance (for example, when the surrounding space is
filled by a dielectric with loss), an impedance-matched,
non-reflecting body --- that is the black body in its conventional and
intuitive definition --- will not anymore be the one that attains the
maximal spectral emissivity.

Let us also mention one important case when the impedance matching
condition is {\em sufficient} to maximize the power emitted from a
body to free space. It is the case when the emitting body is so large
that it can be modeled with the geometry of Fig.~\ref{circuit}(b).  As
was mentioned in Sec.~\ref{sectmodel}, in this case the basis of
orthogonal spatial harmonics is composed of propagating and evanescent
plane waves. The wave impedances of these modes are given by
Eq.~\r{zwave}. The evanescent plane waves with $q > k_0$ have
$\Re\left(Z_w^{\rm TE,TM}\right) = 0$ and, thus, in accordance with
Eq.~\r{specdensbb}, do not contribute into the far-field emission at
all. The propagating modes with $q<k_0$ have $\Re\left(Z_w^{\rm
    TE,TM}\right) > 0$ and $\Im\left(Z_w^{\rm TE,TM}\right) =
0$. Because the wave impedance of these modes is purely real, the
maximum emission condition $Z_1^* = Z_2$ for these waves coincides
with the condition $Z_1 = Z_2$. Therefore, the optimal emitter in this
case is the half-space with zero reflection: $\Gamma = 0$, i.e., the
black half-space. This essentially forbids any far-field
super-Planckian emission in such geometries. However, it does not
follow from here that the same conclusion must hold in geometries
involving objects of finite size.

\section{Free space far-field thermal emission from bodies of finite size}
\label{cmemitter}

Let us now focus on geometries involving optically large spherical
emitters in free space, or, more generally, any finite size emitters
which completely fit into a sphere of a fixed radius $r = a\gg
\lambda$. An analogous treatment can be applied to cylindrical
emitters.

It is well-known that the electromagnetic field produced by
the sources that are fully contained within a finite volume can be
expanded (in the space outside this volume) over the complete set of
vectorial spherical waves, defined with respect to a spherical
coordinate system $(r,\theta,\varphi)$ whose origin lies within this
volume. These modes split into TE waves (with $E_r = 0$) and TM waves
(with $H_r =0$), with the field vectors expressed through a pair of
scalar potentials $U_{lm}, V_{lm} \propto {\cal
  R}_l(k_0r)Y_l^m(\theta,\varphi)$, where $Y_l^m(\theta,\varphi)$ are
Laplace's spherical harmonics, and ${\cal R}_l(x) = xh^{(1)}_l(x)$
with $h^{(1)}_l(x)$ being the spherical Hankel function of the first
kind and order~$l$. The function ${\cal R}_l(x)$ is also known as the
Riccati-Hankel function of the first kind. The transverse electric
field $\_E_{\rm t} = E_\theta\hat{\boldsymbol\thetaup} +
E_\varphi\hat{\boldsymbol\varphiup}$ and the transverse magnetic field
$\_H_{\rm t} = H_\theta\hat{\boldsymbol\thetaup} +
H_\varphi\hat{\boldsymbol\varphiup}$ in these modes are related as
$\_E_{{\rm t},lm} = -Z_{w,lm}^{{\rm TE,TM}}(\hat{\_r}\x\_H_{{\rm
    t},lm})$ (see~\ref{AppA}), where $Z_{w,lm}^{{\rm TE,TM}}$ is the
wave impedance of the spherical wave harmonic with the polar index $l$
($l = 1, 2, \dots$) and the azimuthal index $m$ ($m = 0, \pm 1, \pm 2,
\ldots \pm l$), which can be expressed as
\[
Z_{w,lm}^{{\rm TE}} = i\eta_0{{\cal R}_l(k_0r)\over {\cal
    R}_l'(k_0r)}, \quad
Z_{w,lm}^{{\rm TM}} = -i\eta_0{{\cal R}_l'(k_0r)\over {\cal R}_l(k_0r)}.
\l{spimp}
\]
Note that the wave impedance of a mode depends on the radial distance
$r$ and the polar index $l$, and it is independent of the azimuthal
index $m$. We exclude the purely longitudinal mode with $l = m = 0$
because it does not contribute into the radiated power.

Wave impedances of spatial harmonics~\r{spimp} correspond to spherical
waves emitted from an object comprising the point $r = 0$. For
incoming spherical waves (see, e.g., Ref.~\cite{Nisbet}), the
wave impedances are expressed through the Riccati-Hankel functions of
the second kind $\tilde{{\cal R}}_l(x) = xh^{(2)}_l(x)$, $x = k_0r$,
and are equal to the complex conjugate of the impedances given by
Eq.~\r{spimp}. Such waves are also called anti-causal waves, because
they cannot be created just by remote external sources: A presence of
a scatterer (which is sometimes called ``sink'', as opposed to
``source''~\cite{Nisbet}) in the vicinity of point $r = 0$ is necessary for them to
appear. Nevertheless, it is convenient to use such waves to describe
the heat transfer {\em from} the remote environment {\em to} the body
surface (we will use such waves in Sec.~\ref{scattering}
and~\ref{kirchhoff}).

The striking difference in the properties of the spherical wave
harmonics as compared to the plane wave harmonics discussed in the
last paragraph of Sec.~\ref{sectconj}, is that the wave
impedance~\r{spimp} has a non-vanishing real part
$\Re\left(Z_{w,lm}^{{\rm TE,TM}}\right) >0$ for the harmonics with
arbitrary high indices $l$ and $m$. Therefore, there are no fully
evanescent waves among the spherical wave harmonics: Each mode,
whatever high index it has, contributes into the far field. Hence,
based on the results of Sec.~\ref{sectconj}, we may conclude that at
any given wavelength there is a possibility to satisfy conjugate matching
condition for the {\em entire spectrum} of spherical waves that are emitted
by a body with a finite radius. In this case, a special emitter must
be crafted which will provide the input impedance $Z_1 =
\left(Z_{w,lm}^{\rm TE,TM}\right)^*$ for all the modes with arbitrary
indices $l$ and $m$. We postpone the discussion of this realization
until Sec.~\ref{realization}.

Under such perfect conjugate matching condition, the total power (per
unit of frequency) emitted by the body at a given wavelength (with
both TE and TM polarizations taken into account) satisfy
\[ {dP_{\rm tot}\over d\nu} =2\times\sum_{l=1}^\infty\sum_{m=-l}^l \Theta(\nu, T_1)\rightarrow \infty,
\l{totpow}
\]
i.e., it grows infinitely. Thus, at least from a purely theoretical
point of view, there is no upper limit on the power spectral density
of the far-zone thermal radiation produced by a body of a constrained
radius at a given wavelength. Note that in this consideration we did
not make any assumptions regarding the radius-to-wavelength ratio or
the internal structure of the body.

\begin{figure}[tb]
\centering
\epsfig{file=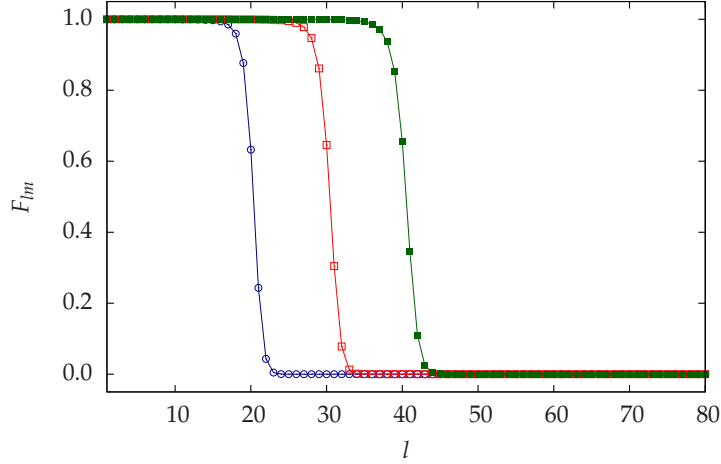,width=0.6\textwidth}
\caption{\label{plotF} Emissivity factor $F_{lm}$ as a
  function of the spherical harmonic polar index $l$ for emitters with
  normalized radii $k_0a = 20$ (blue line with empty circles),
  $k_0a=30$ (red line with empty squares), and $k_0a=40$ (green line
  with filled squares). The value of this factor is the same for TE
  and TM-polarized waves.}
\end{figure}

In order to understand this result, let us compare the perfect
conjugate-matched case of Eq.~\r{totpow} with the case when the
emitter is simply impedance matched, which is expressed in our
equivalent circuit model by the condition $Z_1 = Z_2\equiv
Z_{w,lm}^{\rm TE,TM}$. Under this condition, the power spectral
density per a single spherical harmonic can be expressed from
Eq.~\r{specdensbb}, which leads to
\[
{dP_{\rm tot}\over d\nu} =
\sum_{p=\rm TE,TM}\sum_{l=1}^{\infty}\sum_{m=-l}^l
\left({\Re\left(Z_{w,lm}^{p}\right)\over\left|Z_{w,lm}^{p}\right|}\right)^2\Theta(\nu,T_1),
\l{powbb}
\]
where the index $p = \rm TE,TM$ labels the polarization.
The factor $F_{lm} = {\Re\left(Z_{w,lm}^{\rm
      TE,TM}\right)^2\big/\left|Z_{w,lm}^{\rm TE,TM}\right|^2}$ on the
right-hand side of Eq.~\r{powbb} is close to unity for $l \lesssim
N_{\rm max} = 2\pi a/\lambda$ and decreases to zero very rapidly when
$l > N_{\rm max}$. This is illustrated in Fig.~\ref{plotF}.  The
reason for this is that in the spherical waves with $l > N_{\rm max}$
the electromagnetic energy in the vicinity of emitter's surface is
mostly concentrated in the near fields (reactive fields) which decay
faster than $1/r$ with distance and, thus, do not contribute into the
far field. Respectively, the wave impedance of these waves is such
that $\Re\left(Z_{w,lm}^{\rm TE,TM}\right) \ll
\left|\Im\left(Z_{w,lm}^{\rm TE,TM}\right)\right|$, which results in
the emissivity cut-off at about $l \approx N_{\rm max}$. The same
cut-off can be explained also by the fact that on the surface $r = a$,
a spherical wave harmonic with an index $l\gg 1$ forms a wave pattern
with the characteristic spatial period $t \approx 2\pi
a/l$. Therefore, when $l > 2\pi a/\lambda$, this period is less than
the wavelength so that the mentioned mode behaves at the surface $r = a$
similarly to an evanescent plane wave. Note that such a cut-off is not
present under the conjugate matching condition $Z_1=Z_2^*$, because in
this case the reactive components in $Z_1$ and $Z_2$ have opposite
signs and compensate one another, i.e., the conjugate matching condition
is essentially a resonant condition in the equivalent circuit of
Fig.~\ref{circuit}(a).

Therefore, when dealing with an impedance matched emitter, we may
approximate $F_{lm}$ by unity when $l \le N_{\rm max}$, and by zero
when $l > N_{\rm max}$ when $N_{\rm max} = 2\pi a/\lambda \gg 1$. This
excludes the modes with $l > N_{\rm max}$ from summation~\r{powbb},
and we obtain the following closed-form expression for the power
spectral density:
\[
{dP_{\rm tot}\over d\nu} = 2\times\sum_{l=1}^{N_{\rm max}}\sum_{m=-l}^l
\Theta(\nu,T_1)
= 2N_{\rm max}(N_{\rm max} + 2)\,{h\nu\over \ds\exp{h\nu\over k_{\rm
      B}T_1}-1},
\]
From this formula, by substituting $N_{\rm max} = 2\pi a/\lambda$ and
taking into account that $N_{\rm max} \gg 1$, we get
\[
{dP_{\rm tot}\over d\nu} = {8\pi^2 a^2\over\lambda^2}{h\nu\over \ds \exp{h\nu\over k_{\rm
      B}T_1}-1}.
\l{pow2}
\]
Recognizing $4\pi a^2$ as the area of the spherical surface, we obtain
from Eq.~\r{pow2}
\[ {d^2P_{\rm tot}\over d\nu\,dS} = {2\pi h\nu^3\over c^2}
  {1\over\ds\exp{h\nu\over k_{\rm B}T_1}-1},
\]
which coincides with the amount of power (per unit frequency and
unit area) emitted by a black body sphere kept under temperature
$T=T_1$: ${d^2P/(d\nu\,dS)} = \pi B_\nu(T)$, where
$B_\nu(T)$ is Planck's black body spectral radiance.

Thus, an optically large impedance matched body is equivalent in its
emissive properties to Planck's black body. This is expected, because
such a body is typically understood as made of a black, non-reflecting
material with impedance matched to the free-space impedance at
arbitrary angles of incidence. Black bodies of finite size modeled as
apertures in walls of large opaque cavities also behave similarly,
due to the full match between the domains inside and outside the
cavity.

It is instructive to relate the number of independent spherical
harmonics into which a hot body can emit with the number of photonic
states in free space. Let a spherical body with radius $a \gg \lambda$
be situated in vacuum, and consider a free space gap with thickness
$h$ (an empty spherical layer) adjacent to it such that $\lambda \ll h
\ll a$. Then, the number of photonic states within this gap which
transfer energy away from the body is
\[
dN_{\rm ph} \approx {1\over 2}\x (4\pi a^2) \x h \x
D_{\nu}^{(3)}\,d\nu,
\l{dN1}
\]
where $D_{\nu}^{(3)} = 8\pi\nu^2/c^3$
is the photonic density of states in vacuum.

The same quantity can be also expressed by counting independent
spherical waves within the same gap:
\[
dN_{\rm ph} \approx {1\over 2} \x 2N_{\rm max}(N_{\rm max} + 2) \x h
\x D_{\nu}^{(1)}\,d\nu,
\l{dN2}
\]
where $N_{\rm max} \gg 1$ is the maximal spherical harmonic index up
to which the body emits efficiently, and $D_{\nu}^{(1)} = 4/c$ is the
one-dimensional photonic density of states.

By comparing Eqs.~\r{dN1} and~\r{dN2} we find that $N_{\rm max}\approx
2\pi a/\lambda$, i.e., the same limit as for a black body. However,
note that, by definition, the photonic density of states in vacuum
accounts only for the states which correspond to {\em propagating}
waves, i.e., to real photons. Evanescent waves --- which correspond to
virtual, tunneling photons --- are not accounted in such
description. Respectively, emission above the black body limit is
possible only by such tunneling. For instance, radiating in a mode
with index $l = \kappa N_{\rm max}$, where $\kappa > 1$, must involve
photon tunneling from the body surface at $r = a$ to the distant
surface $r = \kappa a$, where there is an available photonic state for
it. Due to this process, the radiative heat flux will be
super-Planckian in the range of radial distances $a<r<\kappa a$. At
distances greater than $\kappa a$, the radiation flux will remain
sub-Planckian in the sense that its spectral density does not exceed
the one produced by Kirchhoff-Planck's black body with radius $r =
\kappa a$.

Thus, the conjugate-matched body can emit significantly higher power
per unit of area and per unit of frequency as compared to Planck's
black body of the same size only because the conjugate matching
condition tunes the emitter at resonance with high-order modes, due to
which these modes are excited with a very high amplitude. Although
these modes are essentially {\em dark modes} (because they are very
weakly coupled to free space), the resonance greatly increases
probability of photon tunneling from one of such states at emitter's
surface to a propagating free space state at some large enough radial
distance.  Obviously, such a resonant photon tunneling effect is not
possible with a body made of a simple absorbing material.

Finally, let us note that the conjugate matching condition at an
emitting spherical surface mathematically coincides with the
zero-reflectance condition for the anti-causal spherical waves
incident on the same surface (see Sec.~\ref{scattering} for more
detail). In such picture, the conjugate-matched emitter stands out as
a perfect sink for these waves. The energy transferred by
such waves is totally absorbed at emitter's surface without any
reflections. Thus, for finite size emitters, one might try to amend
the definition of the ideal black body in such a manner that it would
refer to the conjugate-matched emitter rather than to the impedance
matched one. One, however, would have to accept in this case that such
a redefined ideal black body is characterized by infinite effective
absorption cross section independently of its real, physical
size. This point is discussed with more detail in the next section.

\section{\label{scattering}Scattering, absorption, and extinction cross sections of
  finite size bodies under conjugate matching condition}

The scattering, absorption, and extinction cross sections at a given
frequency are, by definition,
\[
\sigma_{\rm sc} = {P_{\rm sc}\over \Pi_{\rm inc}}, \quad \sigma_{\rm abs} = {P_{\rm abs}\over \Pi_{\rm inc}}, \quad
\sigma_{\rm ext} = {P_{\rm ext}\over \Pi_{\rm inc}},
\]
where $\Pi_{\rm inc} = {\eta_0^{-1}}|E_{\rm inc}|^2$ is the power flow
density in an incident plane wave with the given frequency, and
$P_{\rm sc}$, $P_{\rm abs}$, and $P_{\rm ext} = P_{\rm sc} + P_{\rm
  abs}$ are, respectively, the amounts of power scattered by the body,
absorbed within it, and extracted by it from the incident
field. Without any loss in generality, we may assume that the incident
wave is propagating along the $z$-axis, and is linearly polarized:
$\_E^{\rm inc} = \hat{\_x}E_{\rm inc}\exp(ik_0z)$.

As is well known, an incident plane wave can be expanded into
vectorial spherical waves. The notations for such waves vary a lot in
literature, but when reduced down to Riccati-Bessel functions and derivatives
of the Laplace spherical harmonics the expansion can be written as
follows (only the electric field component transverse to $\hat{\_r} = \_r/|\_r|$ is of
our interest):
\begin{eqnarray}
\_E^{\rm inc}_{\rm t} &=& -\hat{\_r}\x(\hat{\_r}\x\_E^{\rm inc})=
{\sqrt{\pi}E_{\rm inc}\over k_0r}\sum_{l=1}^\infty\sum_{m=-1,1}\!\!\sqrt{2l+1\over
  l(l+1)}\nonumber\\
&\x& i^{l-1}\left[ {\cal S}_l(k_0r)\,\_r\x\nabla_{\rm t}-m{\cal S}'_l(k_0r)\,r\nabla_{\rm t} \right]Y_l^m(\theta,\varphi),
\l{planewaveexp}
\end{eqnarray}
were ${\cal S}_l(x) = x j_l(x)$ is the Riccati-Bessel function of the
first kind with $j_l(x)$ being the spherical Bessel function of the
first kind and order $l$, and $\nabla_{\rm t} = \nabla - \hat{\_r}(\d/\d
r)$. In the inner summation over $m$ the index acquires just two
values: $m=-1$ and $m=1$. The first term in the square brackets of
Eq.~\r{planewaveexp} proportional to ${\cal S}_l(k_0r)$ is due to the
TE-polarized spherical waves and the second one proportional to the
derivative of the same function is due to the TM-polarized part of the
spectrum. These two contributions are mutually orthogonal.

Note that the Riccati-Bessel functions in expansion~\r{planewaveexp} are
simple superpositions of the Riccati-Hankel functions ${\cal R}_l(x)$ and $\tilde{\cal R}_l(x)$
(of course, the same refers to their derivatives):
\[
{\cal S}_l(x) = {{\cal R}_l(x)+\tilde{\cal R}_l(x)\over 2}, \quad
{\cal S}'_l(x) = {{\cal R}'_l(x)+\tilde{\cal R}'_l(x)\over 2}.
\]
Therefore, in the plane wave expansion~\r{planewaveexp} there are two types
of spherical waves --- leaving waves propagating towards $r = \infty$ and
incoming ones propagating towards $r = 0$, having exactly the same magnitudes. Thus, the incident wave
expansion~\r{planewaveexp} is essentially an expansion over {\em
  standing} spherical waves. This is not surprising, as the net power
flow through any closed surface (with no enclosed scatterers!) vanishes
for any plane wave. This explains our earlier remarks regarding the
anti-causality of incoming waves. So, these waves may not be excited exclusively by remote sources
alone. Even though such waves seem to arrive from $r = \infty$, they
may be separated from their causal pair {only} due to
scattering on an object located in the vicinity of the point $r = 0$.
The scattering destroys the perfect balance between leaving and incoming
waves which exists in the incident field.

Hence, the expansion~\r{planewaveexp} can be seen as composed of the
counter-propagating TE- and TM-polarized spherical waves with indices $l=1, 2, \ldots$ and
$m = -1, 1$, and with complex amplitudes
\[
A^{\rm TE}_{l,\pm 1} = B^{\rm TE}_{l,\pm 1} = \mp A^{\rm TM}_{l,\pm 1} = \mp B^{\rm TM}_{l,\pm 1} =
i^{l-1}{\sqrt{\pi}E_{\rm inc}\over 2}\!\sqrt{2l+1\over l(l+1)},
\]
where we use letter $A$ to denote waves propagating towards $r = 0$,
and letter $B$ for the oppositely propagating ones. In these notations,
\begin{eqnarray}
\_E^{\rm inc}_{\rm t} &=&
{1\over k_0r}\sum_{l=1}^\infty\sum_{m=-1,1}\left[ A^{\rm TE}_{l,m}\tilde{\cal R}_l(k_0r)\,\_r\x\nabla_{\rm t} +
        A^{\rm TM}_{l,m}\tilde{\cal R}'_l(k_0r)\,r\nabla_{\rm t} +\right.\nonumber\\
& & \quad\quad\quad\quad\left.\ B^{\rm TE}_{l,m}{\cal R}_l(k_0r)\,\_r\x\nabla_{\rm t} +
         B^{\rm TM}_{l,m}{\cal R}'_l(k_0r)\,r\nabla_{\rm t} \right]Y_l^m(\theta,\varphi).
\l{morewaveexp}
\end{eqnarray}

An object located in the vicinity of the point $r = 0$ perturbs the
balance of the incoming and the outgoing spherical waves, resulting in
nonvanishing $P_{\rm sc}$ and $P_{\rm abs}$, and $\sigma_{\rm sc, abs}
\neq 0$. The closed-form expressions for these quantities are derived
in~\ref{AppB}. Although similar derivations can be found in
many sources on optical scattering, in~\ref{AppB} we use our
original impedance-based formalism, which allows us to relate the
scattering theory results to the results of our equivalent circuit
model in the most natural manner.

The normalized scattering cross section is obtained based on the
results of~\ref{AppB} and reads
\[
{\sigma_{\rm sc}\over\pi a^2} = {1\over 4(k_0a)^2}\sum_{p=\rm TE, TM}\sum_{l=1}^\infty\sum_{m=-1,1}(2l+1)
\left|1 - \tilde{\Gamma}_{lm}^{\,p}\right|^2,
\l{sigmasc}
\]
where the reflection coefficients $\tilde{\Gamma}_{lm}^{\rm TE,TM}$
are given by Eq.~\r{gammatildeTETM}.

For an ideally conjugate-matched body, $Z_{1,lm}^{\rm TE,TM} =
\left(Z_{w,lm}^{\rm TE,TM}\right)^*$ [here, $Z_{1,lm}^{\rm TE,TM}$ has
absolutely the same meaning as $Z_1$ in the equivalent circuit of
Fig.~\ref{circuit}(a)], therefore, as it follows from
Eqs.~\r{gammatildeTETM} and~\r{reflTETM}, $\tilde{\Gamma}_{lm}^{\,p} =
0$. Hence, in this case
\[
{\sigma_{\rm sc}\over\pi a^2} = {1\over 4(k_0a)^2}\sum_{p=\rm TE,
  TM}\sum_{l=1}^\infty\sum_{m=-1,1}(2l+1) \rightarrow \infty.
\]
Note that our derivation remains valid when $k_0a \gg 1$, thus, we may
conclude that even for optically large bodies, the scattering cross
section is not limited by the geometric cross section and can be
arbitrary high. Moreover, because the total power associated with an
incident plane wave is infinite, the power scattered by an object can
also be arbitrary high.

Our model allows us to conclude also that an optically large body with
radius $r = a$ made of an absorbing material with characteristic
impedance close to that of free space will behave similarly to the
impedance matched body considered in Sec.~\ref{cmemitter}. Namely, for
such a body, $\tilde{\Gamma}_{lm}^{\,p}\approx 0$ for modes with $l \le
N_{\rm max} = 2\pi a/\lambda$, $N_{\rm max} \gg 1$, and
$\tilde{\Gamma}_{lm}^{\,p}\approx 1$ for the modes with
$l > N_{\rm max}$. Respectively, the normalized scattering
cross section of such a body is
\[
{\sigma_{\rm sc}\over\pi a^2} \approx {1\over 4(k_0a)^2}\sum_{p=\rm TE,
  TM}\sum_{l=1}^{N_{\rm max}}\sum_{m=-1,1}(2l+1) {N_{\rm max}(N_{\rm
    max}+2)\over (k_0a)^2} \approx 1,
\]
i.e., its scattering cross section coincides with the geometric
cross section.

The absorption cross section is derived in~\ref{AppB} and
satisfies
\begin{eqnarray}
{\sigma_{\rm abs}\over \pi a^2} &=& {1\over 4(k_0a)^2}\sum_{p=\rm TE,
  TM}\sum_{l=1}^\infty\sum_{m=-1,1}(2l+1) \left(1 -
    \left|\tilde{\Gamma}_{lm}^{\,p}\right|^2\right)\nonumber\\
&=&{1\over (k_0a)^2}\sum_{p=\rm TE, TM}\sum_{l=1}^\infty\sum_{m=-1,1}
(2l+1){\Re\left(Z_{1,lm}^{\,p}\right)\Re\left({Z_{w,lm}^{\,p}}\right)\over
 \big|Z_{1,lm}^{\,p}+Z_{w,lm}^{\,p}\big|^2}.
\l{sigmaabs}
\end{eqnarray}
Note the apparent similarity of the terms under summation~\r{sigmaabs}
with Eq.~\r{pow}. From Eq.~\r{sigmaabs} one can see that because a
perfectly conjugate- matched body is characterized with
$\tilde{\Gamma}_{lm}^{\rm TE,TM} = 0$, the absorption cross section of
it is infinite, similarly to the scattering cross section we have
found earlier. It is also directly seen that the conjugate matching
condition maximizes the summation terms in
Eq.~\r{sigmaabs}. Analogously to what have been done earlier, one may
verify that the absorption cross section of a large impedance matched
body is $\sigma_{\rm abs} = \sigma_{\rm sc} = \pi a^2$. Finally, the
extinction cross section of an arbitrary body can be found from
Eq.~\r{sigmasc} and \r{sigmaabs} as $\sigma_{\rm ext} = \sigma_{\rm
  sc} + \sigma_{\rm abs}$.

\begin{figure}[tb]
\centering
\epsfig{file=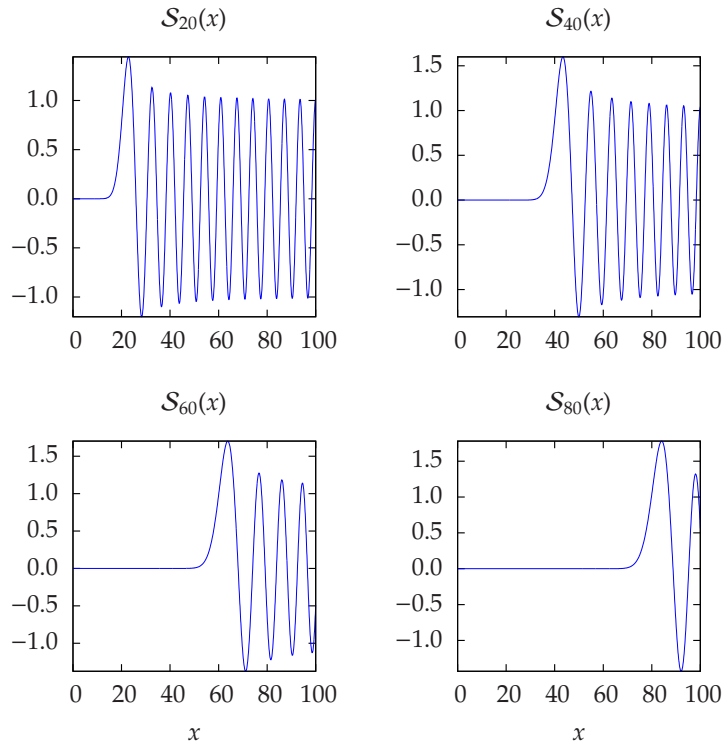,width=0.6\textwidth}
\caption{\label{standwave} Radial dependence functions ${\cal S}_l(x)$
  for $l=20$, 40, 60, and 80. Note that these functions decay quickly
  to zero for $x < l$, when $x$ approaches zero.}
\end{figure}

To conclude with the study of this section let us try to analyze (now
from the point of view of scattering theory) what property makes it
possible to achieve the values of the normalized absorption cross
sections much greater than unity, which, reciprocally, increases in
the same proportion the effective spectral emissivity of a body. In order to do
this, let us consider the behavior of the incident field
expansion~\r{planewaveexp} within the spherical domain $r \le
a$. Although {\em all} terms in expansion~\r{planewaveexp} produce a
non-vanishing contribution within this region, {\em the dominant}
contribution is due to the modes with polar indices $l$ such that $l
\lesssim k_0a = 2\pi a/\lambda$. Mathematically, this can be seen from
the asymptotic behavior of Riccati-Bessel functions ${\cal S}_l(x)$ at
small values of argument: ${\cal S}_l(x) \sim
(x/2)^{l+1}[\!\sqrt{\pi}/\Gamma(l+3/2)]$ (here, $\Gamma(z)$ is the
Gamma-function; this asymptotic is valid up to $x^2 \lesssim l$),
therefore, the modes with $l \gg k_0a$ quickly decay in this region
when $x = k_0r$ approaches zero. Fig.~\ref{standwave} shows the radial
behavior of a few spherical standing waves in the vicinity of this
region.

This demonstrates that when an object such as, for example, a ball
made of some absorbing material is placed in the region $r < a$ it
will mostly interact with the modes with the indices $l \lesssim
k_0a$. Thus, the power transport from the remote environment to this
object will be mediated by these modes dominantly. Reciprocally, when
the object is the source of thermal radiation, the fluctuating
currents in the object will excite the same set of modes, so that only
the modes with $l \lesssim k_0a$ will participate in the reversely
directed heat transport. However, it is not hard to imagine that a
specially crafted body can be forced to interact also with the higher
order modes with $l \gg k_0a$, because, besides being weak, these
modes nevertheless penetrate into the region $r < a$. The strongest
interaction is achieved at the resonant condition $Z_{1,\,lm}^{\rm
  TE,TM} = \left(Z_{w,\,lm}^{\rm TE,TM}\right)^*$, which maximizes the
terms of summation~\r{sigmaabs}. Thus, the physical reason for the
increased interaction is this resonance.

\section{\label{kirchhoff}Implications with regard to second law of thermodynamics and
  Kirchhoff's law of thermal radiation}

One may think that the result of Sec.~\ref{cmemitter} ---
which essentially states that a body of an optically large but finite
size may emit, theoretically, arbitrarily high power per unit of
frequency and per unit of area --- contradicts the second law of
thermodynamics. For instance, earlier claims of super-Planckian
thermal radiation from photonic crystals were rebutted in
Ref.~\cite{2nd_law} on this ground (we discuss this reference in
more detail later in this section). This is, however, not our
case. From the equivalent circuit of Fig.~\ref{circuit}(a) it is
immediately understood that the conjugate matching condition that
maximizes the radiated power, at the same time maximizes the power
delivered from the environment back to the emitting body, i.e., the
optimal heat emitter is at the same time the optimal heat sink! The
same conclusion can be drawn from the results Sec.~\ref{scattering},
in which we have demonstrated that the absorption cross section of a
body is maximized under the same conjugate matching condition.
Therefore, the conjugate matching condition preserves the balance of
radiative heat exchange between the body and its environment when $T_1
= T_2$.  The symmetry of the equivalent circuit (a consequence of the
reciprocity principle), actually, simply forbids obtaining from our
theory any result that would violate such thermodynamical heat
exchange balance. On the other hand, from the same circuit it
immediately follows that when $T_1 \neq T_2$ the net radiative heat
flow is always directed from the side with higher temperature to the
side with lower temperature.

In order to study implications of our theory with regard to
Kirchhoff's law let us inspect how large is the power $dP_{\rm inc}$
associated with an incoming spherical wave incident from the the side
of the remote environment (kept at temperature $T_2$), in a general
scenario when the input impedance of the body $Z_1\neq
\left(Z_{w,lm}^{\rm TE,TM}\right)^*$. The environment --- free space
in our case --- is characterized by impedance $Z_2 = Z_{w,lm}^{\rm
  TE,TM}$.

The expression for the power $dP_{21}$ received by the body from the
environment is readily obtained from the equivalent circuit model:
\[
dP_{\rm 21} = {4\Re(Z_1)\Re(Z_2)\over\left|Z_1+Z_2\right|^2}\,\Theta(\nu, T_2)\,d\nu.
\l{pow21}
\]
This power can be split into the
incident and reflected power: $dP_{21} = dP_{\rm inc} - dP_{\rm ref}$,
where
\[
dP_{\rm ref} = \tilde{\rho}\,dP_{\rm inc} =  \left|{Z_1-Z_2^*\over Z_1 + Z_2}\right|^2dP_{\rm inc}.
\l{powref}
\]
In this formula, $Z_2^* = \left(Z_{w,lm}^{\rm TE,TM}\right)^*$ is the
wave impedance for the spherical wave propagating from $r
= \infty$ towards $r = 0$, and $\tilde{\rho} = \left|{Z_1-Z_2^*\over
    Z_1 + Z_2}\right|^2$ is the power reflection coefficient of this
wave at the body surface. Thus, by
combining Eqs.~\r{pow21} and~\r{powref}, we obtain
\[
dP_{\rm inc} = {4\Re(Z_1)\Re(Z_2)\over\left|Z_1+Z_2\right|^2}
{\Theta(\nu, T_2)\,d\nu\over 1 - \left|{Z_1-Z_2^*\over Z_1 +
      Z_2}\right|^2} = \Theta(\nu, T_2)\,d\nu.  \l{powinc}
\]

The above result shows that in a free space environment filled with
thermal-fluctuating electromagnetic field characterized with
temperature $T_2 > 0$ {\em every} spherical harmonic propagating
towards the point $r = 0$ delivers the same amount of heat power:
$dP_{\rm inc} = \Theta(\nu, T_2)\,d\nu$. Thus, the amount of power
transported by all such harmonics with arbitrary indices $l$ and $m$
to an object comprising the point $r = 0$ is
infinite. Kirchhoff-Planck's black bodies and ordinary absorbers {\em
  reflect} most of this incident power: Only the power delivered by
the incident modes with $l \le N_{\rm max}$ (see
Sec.~\ref{scattering}) can be efficiently received by such
bodies. Reciprocally, they radiate back only into the same limited
number of outgoing modes. On the contrary, an ideal conjugate-matched
body is theoretically able to receive the whole infinite power
delivered by {\em all} such incident waves, as well as to radiate it
back.

Comparing Eq.~\r{powinc} with Eq.~\r{pow} when $T_1=T_2=T$ we may write
\[
{dP_{12}\over \alpha} = \Theta(\nu, T)\,d\nu,
\l{genkirch}
\]
where $\alpha = 1 - \left|{Z_1-Z_2^*\over Z_1 +
    Z_2}\right|^2$. Eq.~\r{genkirch} is a generalization of
Kirchhoff's law of thermal radiation. Indeed, the dimensionless
parameter $\alpha = 1 - \tilde{\rho}$ has the meaning of absorptivity
of the body for a given spatial harmonic; $dP_{12}$ is the emitted
power at the same spectral component; and the ratio of these two
quantities is a universal function of just frequency and temperature.
Note that unlike the classical law of the same name, Eq.~\r{genkirch}
is written for a single component of the spatial spectrum of the
radiated field. Thus, Eq.~\r{genkirch} complements the principle of
detailed balance by making it applicable to separate spatial harmonics
of the radiated field.

Thus the classical Kirchhoff law of thermal radiation, which states
that: ``For a body of any arbitrary material, emitting and absorbing
thermal electromagnetic radiation at every wavelength in thermodynamic
equilibrium, the ratio of its emissive power to its dimensionless
coefficient of absorption is equal to a universal function only of
radiative wavelength and temperature --- the perfect black body
emissive power,'' will hold for {\em any} emitter --- being optically
small or large, including the ones characterized with $\sigma_{\rm
  abs}$ much greater than their geometric cross section, --- if we
disregard the intuitive definition of the perfect black body as an
``ultimate absorber'' which attains the absolute maximum in
absorptivity (as compared to all other bodies), in favor of defining
the ideal black body just as an {\em abstract object} characterized
with the emissive power~\r{genkirch}.  Indeed, a conjugate-matched
body receives from the environment and absorbs a much greater power
than the conventional black body of the same size, so that its
effective absorptivity relative to the black body is much greater than
unity: $\sigma_{\rm abs}/(\pi a^2) \gg 1$ (see Sec.~\ref{scattering}).

Finally, let us consider a thought experiment described in
Ref.~\cite{2nd_law}, where a presumably super-Planckian thermal
radiator exchanges thermal energy with an ideal black body located in
far zone. If both objects are infinite in spatial extent (two parallel
infinite slabs), super-Planckian far-field radiation is impossible, as
we have proven in Sec.~\ref{sectconj}, and there is no need for
thermodynamic considerations to prove that again. On the other hand, when
a pair of finite-size bodies are separated by a
distance significantly greater than
$d = \sqrt{\max(\sigma_{\rm abs,1}, \sigma_{\rm abs,2})/\pi}$
the heat exchange flux is also sub-Planckian.

Next, by considering the case of two finite-size bodies separated by a
distance smaller than $d$ but still much greater than the wavelength,
we note that consideration from Ref.~\cite{2nd_law} would assume
in this case that the black body perfectly absorbs {\em all} incident
power, while its radiation is, naturally, restricted by the Planck law
of black body radiation. This assumption, obviously, implies a
violation of the second law of thermodynamics, because it violates the
reciprocity of the heat exchange.

However, as we have shown above, Kirchhoff-Planck's black body
perfectly absorbs only the fully propagating part (ray part) of the
incident spatial spectrum. The higher-order spherical harmonics
incident on the black body surface which are responsible for the
super-Planckian part of the radiative heat, will be reflected from its
surface, and only the ray part will be ideally absorbed and re-emitted
by the receiver. The reflected super-Planckian part of the radiation
will be scattered into the surrounding space. Part of this energy will
then be re-absorbed by the conjugate-matched emitter, which acts as
the ideal sink for all incident waves.

Obviously, when the two bodies have the same temperature, there is no
net heat flow between them, and, thus, the second law of
thermodynamics is not violated. One can say that the Nature avoids
such violation by totally reflecting the photons which cannot be
re-emitted by the receiving body. On the other hand, there is no
limitation for emission of those photons into free space from a
conjugate-matched emitter which interacts resonantly with the entire
infinite spectrum of outgoing spherical waves. Such emission is
possible because for such ideal emitters {\em all} photonic states in
the surrounding empty space are available, by the process of resonant
photon tunneling discussed in Sec.~\ref{cmemitter}.

It is instructive to note here that in the above scenario it is still
possible to realize a super-Planckian heat exchange between an
ordinary body (i.e., a real body with loss, as opposed to the ideal
Kirchhoff-Planck's black body discussed above) and a specially crafted
body which is conjugate matched to the modified environment which
takes into account the presence of the first body, i.e., in this case
the conjugate-matched emitter must be designed so that it maximizes
the probability of photon tunneling between the two bodies. Because
the tunneling process is reciprocal, the second law of thermodynamics
is not violated although the ordinary body will emit above Planck's
limit in this scenario.

\section{\label{realization}Conjugate-matched DNG sphere: metamaterial
  ``thermal black hole''}

Let us now discuss the practical implications of our theoretical
findings. It is clear that realizing conjugate matching condition for
waves with polar index $l > 2\pi a/\lambda$ in a practical thermal
emitter is impossible in emitters formed by homogeneous dielectrics or
magnetics with positive constitutive parameters. For such materials,
the standard arguments of Ref.~\cite{Bohren} apply.

Hence, here we shall investigate if a magneto-dielectric sphere filled
by a material with less restricted parameters $\E$ and $\M$ (e.g.,
{\em a metamaterial}) can be used in realization of the conjugate-matched emitter. It is known that the complex permittivity and
permeability of a passive material at a given frequency can have
either positive or negative real parts, while the sign of the
imaginary part is fixed: $\Im(\E,\M) \ge 0$. Because our goal is to
realize an omnidirectional emitter, the material parameters $\E$ and
$\M$ may depend on $r$, but should not depend on the angles $\varphi$
and $\theta$. Thus, we are ought to find such $\E(r)$ and $\M(r)$ that
will make the input impedance of a sphere made of this material to
become equal to the complex conjugate of impedance~\r{spimp}
for spherical harmonics with arbitrary indices.

In order to do that we first solve an auxiliary problem: With which
{\em uniform} material should we fill the domain $r > a$, so that the
input impedance of {\em this} domain becomes the complex conjugate
of~\r{spimp}? The answer to such a question can be obtained from
Eq.~\r{spimp} generalized for the case when $\E, \M\neq \E_0,\M_0$,
which reads
\begin{eqnarray}
\l{spimpem}
\tilde{Z}_{w,lm}^{{\rm TE}} &=& i\sqrt{\M\over\E}\,{{\cal R}_l(\o\!\sqrt{\E\M}\,r)\over {\cal R}_l'(\o\!\sqrt{\E\M}\,r)},\\
\tilde{Z}_{w,lm}^{{\rm TM}} &=& -i\sqrt{\M\over\E}\,{{\cal R}_l'(\o\!\sqrt{\E\M}\,r)\over {\cal R}_l(\o\!\sqrt{\E\M}\,r)}.
\end{eqnarray}
Consider the properties of the radial function ${\cal R}_l(x)$ and its derivative:
\[
{\cal R}_l(-x) = (-1)^{l+1}\big[{\cal R}_l(x)\big]^*, \quad
{\cal R}'_l(-x) = (-1)^l\big[{\cal R}'_l(x)\big]^*,
\l{rrprop}
\]
which hold when $\Im(x) \rightarrow 0$. When $\E$ and $\M$ are such
that $\E = -\E_0(1-i|\tan\delta|)$, $\M = -\M_0(1-i|\tan\delta|)$ with
loss tangent $|\tan\delta|\rightarrow 0$, the refractive index becomes
$n=\sqrt{\E\M/(\E_0\M_0)}\rightarrow -1$, and we obtain
from~\r{spimpem}--\r{rrprop} for the input impedance of the domain $r
> a$ filled with such material:
\[
Z_{r>a}^{{\rm TE}} \rightarrow -i\eta_0\left({{\cal R}_l(k_0r)\over {\cal R}_l'(k_0r)}\right)^*, \quad
Z_{r>a}^{{\rm TM}} \rightarrow i\eta_0\left({{\cal R}_l'(k_0r)\over {\cal R}_l(k_0r)}\right)^*,
\l{Zrgta}
\]
which is exactly the complex conjugate of the wave impedance~\r{spimp}.

However, we need to obtain such result not for the input impedance of
the domain $r > a$, but for the input impedance of the domain $r < a$
occupied by the emitter. Hence, we need to amend the above
consideration somehow so that it applies to the domain $r < a$. This
can be achieved by applying a proper coordinate transformation to the
Maxwell equations. Such transformation should map the domain $r > a$
into the domain $r < a$ while preserving the field equations in their
usual form. The latter ensures that after such transformation the input
impedance of the transformed domain coincides with the one for the
original domain: $Z_{r<a} = Z_{r>a}$.

The transformation with the required properties is $r \mapsto a^2/r$
derived in~\ref{AppC}. Indeed, under this transformation, $\d/\d
r\mapsto -(r^2/a^2)(\d/\d r)$ and $\hat{\_r}\mapsto
-\hat{\_r}$. Because $\nabla = \nabla_{\rm t} + \hat{\_r}(\d/\d r)$,
where $\nabla_{\rm t} =
(1/r)\left[\hat{\boldsymbol{\theta}}(\d/\d\theta) +
  (\hat{\boldsymbol{\varphi}}/\sin\theta)(\d/\d\varphi)\right]$, the
nabla operator transforms as $\nabla \mapsto (r^2/a^2)\nabla$. This
effectively transforms the material parameters of a uniform
magnetodielectric to $\E\mapsto (a^2/r^2)\E$ and $\M\mapsto
(a^2/r^2)\M$, while preserving the usual form of the Maxwell equations
(see~\ref{AppC} for more detail).

Therefore, a spherical emitter with radius $r = a$ made of a DNG
metamaterial with parameters
\begin{eqnarray}
\l{DNGeps}
\E(r) &=& -{\E_0a^2\over r^2}(1-i|\tan\delta|),\\
\M(r) &=& -{\M_0a^2\over r^2}(1-i|\tan\delta|)
\l{DNGmu}
\end{eqnarray}
will have the input impedance $Z_1 \equiv Z_{r<a}$ coincident with
Eq.~\r{Zrgta} when $|\tan\delta|\rightarrow 0$. Note that this
impedance approaches $\left(Z_{w,lm}^{\rm TE,TM}\right)^*$ arbitrarily
closely when $|\tan\delta|\rightarrow 0$, for modes with arbitrary
indices. Thus, an emitter filled by a material with
parameters~\r{DNGeps} and~\r{DNGmu} constitutes a physical realization
of the conjugate-matched emitter introduced in
Sec.~\ref{cmemitter}.

A similar profile of $|\E(r)| \propto 1/r^2$ was used in the
theoretical~\cite{black_Narimanov} and experimental~\cite{Cheng,
  black_hole_exp} papers where all the materials have positive real
parts of the permittivity and permeability. This leads to a spherical
object which theoretically fully absorbs all rays incident on its
surface, that is, behaves as a Kirchhoff black body. In contrast to
our proposed body whose absorption cross section is theoretically
infinite, the absorption cross section of the ``optical black holes''
described in Refs.~\cite{black_Narimanov,Cheng,black_hole_exp} equals
to the geometric cross section of the body. A quasistatic case in
which a cylindrical body appears having a larger radius than its
physical radius was considered in Ref.~\cite{Zubin}.

Following the same terminology, we may designate a body characterized
with the parameters \r{DNGeps}--\r{DNGmu} as a metamaterial ``black
hole''. With this, we emphasize the property of this object to
intercept rays of light which are not incident directly on its
surface. Due to this feature, such an object has an effective radius
of ray capture (kind of ``Schwartzschild radius'') which can be much
greater than the geometrical radius of the body. The latter, in the
case of the ideal conjugate-matched body, can have arbitrarily small
dimensions, much like the mass singularity in a black hole in
Einstein's gravitation theory. Indeed, as it has been found in
Sec.~\ref{scattering} an ideal conjugate-matched object has infinite
absorption cross section, independently of its geometric size. Hence,
in a geometric optics approximation {\em any} incident ray will end up
hitting such object.

Note however that due to the inevitable dispersion of the DNG medium,
our metamaterial black hole is not ``black'' in the usual optical
sense, as its absorption is frequency dependent.
Our black hole is also different from its astrophysical counterpart in
that sense that when we heat it up it emits light (actually, real
black holes emit radiation and particles as well, but let us not go
too far with such analogies). Because of this, and also because such
an object behaves as an ideal radiative heat sink we may also attach a
label ``thermal'' to its name.

Although, theoretically, $|\tan\delta|$ can be arbitrary small while
still allowing for a non-vanishing loss within the emitter (which is
the necessary condition for thermal emission to occur), in practice,
$|\tan\delta|$ is always finite. Moreover, close to the core of the
emitter the parameters~\r{DNGeps} and~\r{DNGmu} are divergent: $|\E|,
|\M| \rightarrow \infty$, when $r \rightarrow 0$. These factors limit
the number of spherical harmonics of the radiated field for which the
conjugate matching condition can be fulfilled in practice. Therefore, the
situation represented by Eq.~\r{totpow} can never be achieved in an
experiment. One, however, can still expect a significant increase in
the emitted power for emitters with parameters resembling those of
Eqs.~\r{DNGeps} and~\r{DNGmu}, especially, for emitters characterized
with moderate values of the ratio $a/\lambda$. For such emitters we
can have $\sigma_{\rm abs} \gg \pi a^2$, while still
$\sigma_{\rm abs} < \infty$. We may still designate these more
realistic objects as ``black holes'' in our fancy way of giving names,
however, the effective Schwarzschild radius $r_{\rm eff} =
\sqrt{\sigma_{\rm abs}/\pi}$ of these holes will be finite, like for
black holes of any finite mass in astrophysics. Respectively, the
astrophysical counterpart of the ideal conjugate-matched emitter is a
black hole of an infinite mass.

\begin{figure}[tb]
\centering
\epsfig{file=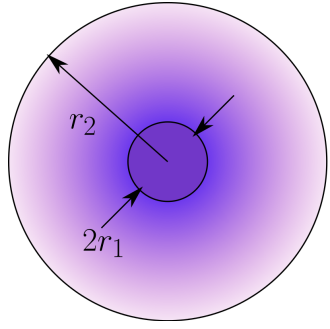,width=0.27\textwidth}
\caption{\label{coreshell} Geometry of the spherical core-shell emitter
  composed of a core with radius $r_1 = a_0$ filled with a uniform
  material with permittivity $\E_{\rm c}$ and permeability $\M_{\rm
    c}$ and the shell with radius $r_2 = a$ filled with a DNG medium
  with radially-dependent parameters $\E(r)$ and $\M(r)$ given by
  Eqs.~\r{DNGeps} and~\r{DNGmu}. The core parameters are matched with
  the shell parameters at $r = a_0$: $\E_{\rm c} = \E(a_0)$, $\M_{\rm
    c} = \M(a_0)$.}
\end{figure}

Let us study the case $\pi a^2 \ll \sigma_{\rm abs} < \infty$ in more
detail and obtain an expression for the effective absorption cross
section of a core-shell metamaterial emitter with parameters $\E(r)$
and $\M(r)$ that follow Eqs.~\r{DNGeps} and~\r{DNGmu} within a
spherical shell $a_0<r<a$, and with uniform parameters $\E_{\rm c} =
\E(a_0)$, $\M_{\rm c} = \M(a_0)$ within the core $r < a_0$ (see
Fig.~\ref{coreshell}). The loss tangent of the DNG metamaterial is
assumed to be finite, but small: $0 < |\tan\delta| \ll 1$.  Note that
there is also another possibility of realizing the necessary level of
absorption in the metamaterial black hole: we may choose $|\tan\delta|
\rightarrow 0$ within the region $a_0 < r < a$, and concentrate all
the loss in the core region $r < a_0$ with a high value of
$|\tan\delta|$.

Let us denote the input impedance of emitter's core for a given
spherical harmonic by $Z_{{\rm c},lm}^{\rm TE, TM}$. This impedance
can be calculated as
\begin{eqnarray}
Z_{{\rm c},lm}^{\rm TE} &=& -i\sqrt{\M_{\rm c}\over\E_{\rm c}}{{\cal
    S}_l(\o\!\sqrt{\E_{\rm c}\M_{\rm c}}a_0)\over {\cal
    S}'_l(\o\!\sqrt{\E_{\rm c}\M_{\rm c}}a_0)},\\
Z_{{\rm c},lm}^{\rm TM} &=& i\sqrt{\M_{\rm c}\over\E_{\rm c}}{{\cal
    S}'_l(\o\!\sqrt{\E_{\rm c}\M_{\rm c}}a_0)\over {\cal
    S}_l(\o\!\sqrt{\E_{\rm c}\M_{\rm c}}a_0)}.
\end{eqnarray}
These relations are analogous to the expressions for the impedance of
the incoming spherical waves, with the radial dependence function
$\tilde{\cal R}_l(x)$ replaced by ${\cal S}_l(x)$. Next, the input
impedance of the whole emitter can be expressed as
\[
Z_{1,lm}^p = Z_{11,lm}^p -
{Z_{12,lm}^pZ_{21,lm}^p\over {Z_{22,lm}+Z_{{\rm c},lm}}},
\l{ZinZparam}
\]
where $Z_{11,lm}^p$, $Z_{12,lm}^p = Z_{21,lm}^p$, and $Z_{22,lm}^p$,
$p = \rm TE,TM$, are the equivalent
$Z$-matrix parameters of the spherical shell $a_0<r<a$ for the same
spherical harmonic. By using the equivalence between the shell region
and the domain $a < r < a^2/a_0$ under the transformation $r \mapsto
a^2/r$, these parameters can be calculated from known formulas for
$Z$-parameters of uniformly filled spherical shells (see~\ref{AppA}).

The absorption cross section can now be calculated with the help of
Eq.~\r{sigmaabs}.  It can be easily checked that for a body with
$\E(r)/\E_0 = \M(r)/\M_0$ (which is our case), $\tilde{\Gamma}_{lm}^{\rm
  TE} = \tilde{\Gamma}_{lm}^{\rm TM} = \tilde{\Gamma}_{lm}$. Also, due to
emitter's symmetry these reflection coefficients are independent of
the azimuthal index $m$. Therefore, from Eq.~\r{sigmaabs} we obtain
the following expression for the absorption cross section:
\[ {\sigma_{\rm abs}\over \pi a^2} = {1\over
  (k_0a)^2}\sum_{l=1}^\infty(2l+1)
{4\Re\left(Z_{1,lm}\right)\Re\left(Z_{w,lm}\right)\over
 \big|Z_{1,lm}+Z_{w,lm}\big|^2},
\l{sigmaabsDNG}
\]
in which the impedances need to be calculated just for a single
polarization (it does not matter for which). The numerical results
obtained with the help of Eq.~\r{sigmaabsDNG} are presented in the next section.

\section{\label{results}Numerical results}

We consider the core-shell emitter depicted in
Fig.~\ref{coreshell}. The shell has the inner radius $r_1 = a_0$ and
the outer radius $r_2 = a$ and is formed by a DNG metamaterial with
parameters given by Eqs.~\r{DNGeps} and~\r{DNGmu}.
The core material
has the parameters $\E_{\rm c} = \E(r_1)$ and $\M_{\rm c} =
\M(r_1)$. As shown in Sec.~\ref{realization}, the normalized
absorption cross section $\sigma_{\rm abs}/(\pi a^2)$ of such a body
must grow without limit when $\tan\delta$ and the ratio $r_1/r_2$ both
tend to zero. In this section we study numerically how fast is this
growth and also identify how large the absorption cross section can
become under typical practical limitations.

\begin{figure}[b!]
\centering
\epsfig{file=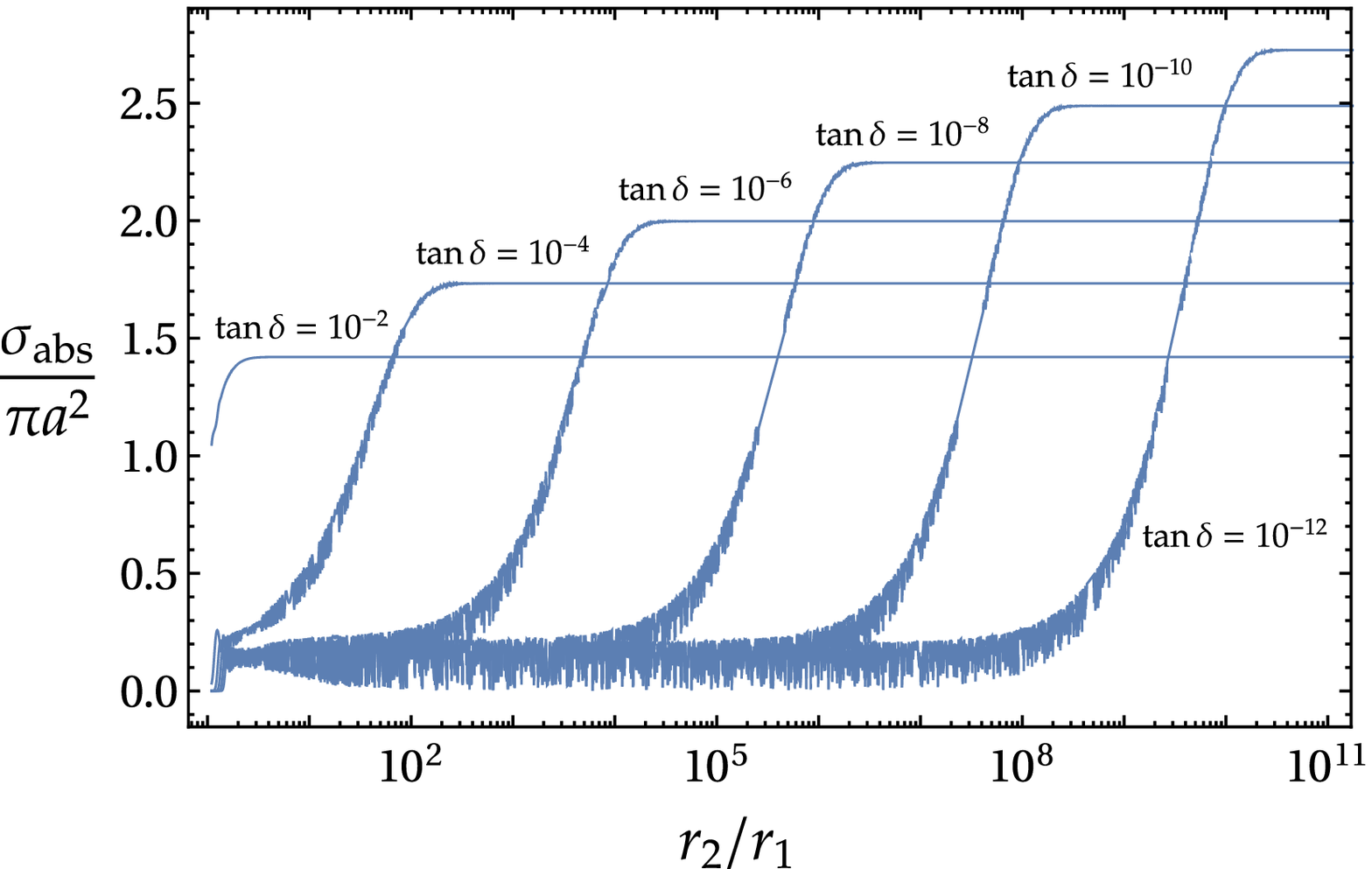,width=0.49\textwidth}
\epsfig{file=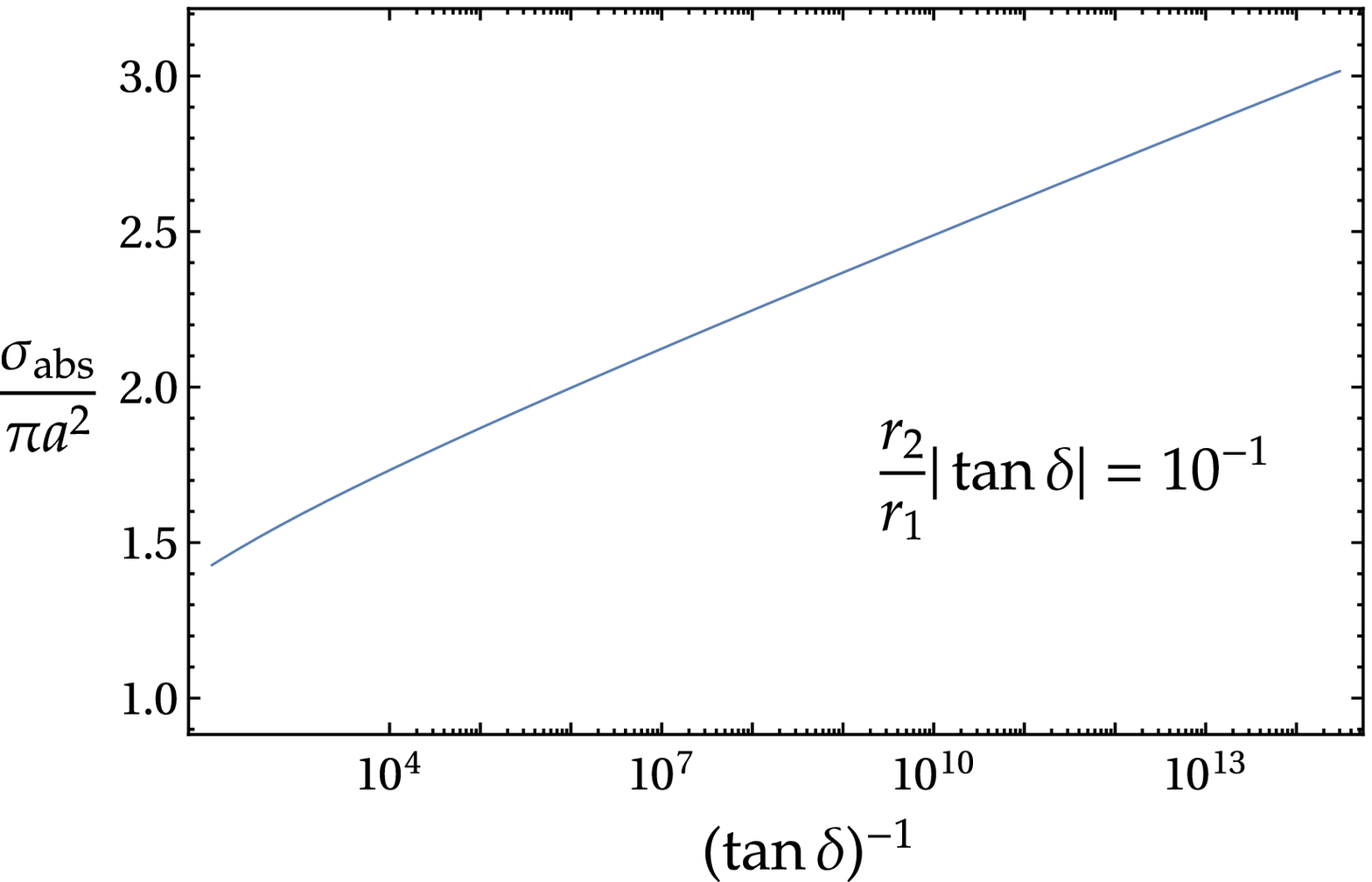,width=0.49\textwidth}
\caption{\label{sigma_core} Left: Normalized absorption cross section
  $\sigma_{\rm abs}/(\pi a^2)$ of the DNG core-shell emitter shown in
  Fig.~\ref{coreshell} as a function of the ratio $r_2/r_1$, for a set
  of fixed values of $|\tan\delta|$ indicated in the plot. Right:
  Normalized absorption cross section as a function of the inverse
  magnitude of $|\tan\delta|$, under the condition that the product
  $(r_2/r_1)|\tan\delta|$ is fixed to the value indicated in the
  plot. In all these numerical examples, $k_0a \equiv k_0r_2 = 30$.}
\end{figure}

For the following it is crucial to note that the limiting behavior of
$\sigma_{\rm abs}$ when $\tan\delta\rightarrow 0$ and $r_1/r_2
\rightarrow 0$ is governed {\em by both these parameters
  together}. Thus different limits can be achieved depending on the
relation between these parameters. For instance, it is trivial to see
that for any fixed ratio $r_1/r_2$,
$\lim\limits_{\tan\delta\rightarrow 0} \sigma_{\rm abs} = 0$, which is
very far from the desired result! However, when $\tan\delta$ is fixed
and the ratio $r_1/r_2$ varies, our numerical analysis shows that the
limiting behavior of $\sigma_{\rm abs}$ changes drastically. For the
core-shell emitter characterized with $k_0a \equiv k_0r_2 = 30$ (which
is optically a relatively large object with circumference of 30
wavelength) this situation is depicted in
Fig.~\ref{sigma_core}~(left).

From this figure one can see that when the core radius $r_1$ decreases
and the ratio $r_2/r_1$ increases, the body absorption cross section
grows at first (with some oscillations related to the thickness
resonances within the core-shell) and later stabilizes at a level
determined by the fixed magnitude of the loss tangent. The tendency is
such that at a smaller $|\tan\delta|$ the achievable $\sigma_{\rm
  abs}$ is higher, but the core radius value required for this becomes
smaller and smaller when the loss tangent decreases.

In this way, arbitrary high values of the normalized absorption cross
section can be achieved provided $\tan\delta$ and the core radius are
decreased {\em together}. For instance, in
Fig.~\ref{sigma_core}~(right) we plot the dependence of $\sigma_{\rm
  abs}$ on the inverse magnitude of $\tan\delta$ under the condition
that the product $(r_2/r_1)|\tan\delta|$ is kept constant. In this
numerical case, when the parameter $|\tan\delta|^{-1}$ grows, the
parameter $r_2/r_1$ increases in the same proportion, which results in
a monotonic growth of the normalized absorption cross section.

However, results depicted in Fig.~\ref{sigma_core} show that 
growth of $\sigma_{\rm abs}/(\pi a^2)$ is very slow. Although
the normalized absorption cross section can theoretically grow without
limit provided suitably small values of the parameters $r_1/r_2$ and
$|\tan\delta|$ are chosen, in practice the range of varying these
parameters is limited.  For instance, considering operating at
wavelengths on the order of 100~$\mu$m, the parameter $r_2/r_1$ can
probably at most reach $10^6$, because higher values of this parameter
would correspond to unrealistically small radii of the inner core (on the order of $1$~nm). As
is seen from Fig.~\ref{sigma_core}~(left) in order to maximize
absorption in this case the loss tangent in the DNG shell must be
decreased down to $10^{-8}$, which is impractical. On the other hand,
attainable loss tangents values $|\tan\delta| > 10^{-4}$, keep the
normalized absorption cross section below 2 (when $k_0a = 30$).

\begin{figure}[tb]
\centering
\epsfig{file=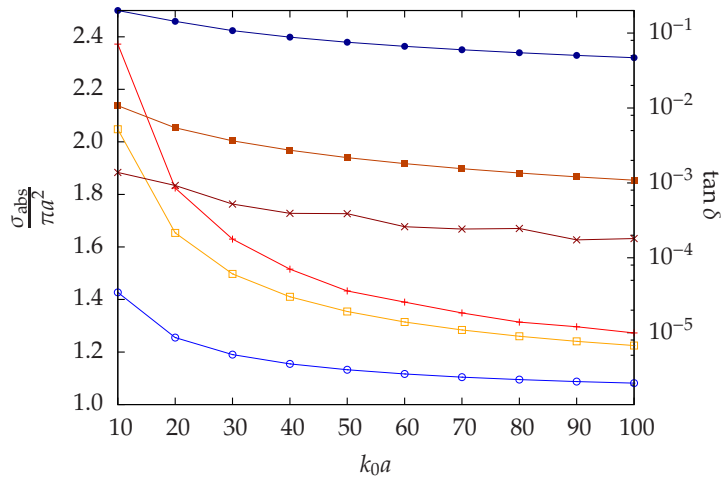,width=0.6\textwidth}
\caption{\label{sigma} Left scale: Normalized
  absorption cross section $\sigma_{\rm abs}/(\pi a^2)$ as a function
  of the normalized radius $k_0a$ of the DNG core-shell emitter shown
  in Fig.~\ref{coreshell}. Red line with ``$+$'' symbols: the case
  when $r_2/r_1 = 100$. Orange line with empty squares: the case
  $r_2/r_1 = 10$. Blue line with empty circles: the case $r_2/r_1 = 1$
  (the case of uniformly filled DNG sphere).
  Right scale: the optimal values of loss tangent $|\tan\delta|$
  (obtained by a numerical optimization procedure) as a function of
  $k_0a$. Dark blue line with filled circles: $r_2/r_1 = 1$. Dark
  orange line will filled squares:  $r_2/r_1 = 10$. Dark red line
  with ``$\x$'' symbols: $r_2/r_1 = 100$. }
\end{figure}

To study these limitations further, we consider the following two
numerical examples.  In the first example, we set $r_2/r_1 = 10$, and,
respectively, $\E(r)$ and $\M(r)$ vary within the shell such that
$\E(r_1)/\E(r_2) = \M(r_1)/\M(r_2) = 10^2$. We
calculate the normalized absorption cross section~\r{sigmaabsDNG} for
a set of $k_0a$ values ranging from $k_0a = 10$ up to $k_0a = 100$
with a step $\Delta k_0a = 10$.  In this range, emitter's
circumference varies in the range from $10\lambda$ up to $100\lambda$,
which indicates that we deal with an optically large body.  The values
of the loss tangent are optimized at each $k_0a$ value (using a
numerical optimization procedure) in order to maximize the normalized
absorption cross section at each point. The result of this calculation
is presented in Fig.~\ref{sigma}.

In the second numerical example we
set $r_2/r_1 = 100$ (i.e., in this case the core is 10 times smaller)
and repeat the same procedure (see Fig.~\ref{sigma}).  We observe that
for $k_0a = 10$ in the first example with $r_2/r_1 = 10$, the
normalized absorption cross section $\sigma_{\rm abs}/(\pi a^2) > 2$,
which means that at this point the DNG core-shell emitter is
performing at least twice better than a black body emitter. Note that
such a result is achieved at a loss level $|\tan\delta| \approx
10^{-2}$ which is significantly less than in typical optical absorbing
materials.

In the literature we could not find any example where the absorption
cross section would be claimed exceeding the geometric one for such a
large object.  In work \cite{Alu2014} targeted to maximization of the
ratio of the absorption cross section to the scattering cross section
it is stressed that the goal is achievable for optically small
particles.  Though the authors of Ref.~\cite{Alu2014} utilize a
similar concept, they combine it with the plasmon resonance of a
core-shell particle. Therefore, their particle with $\sigma_{\rm
  abs}/(\pi a^2)\sim 2\dots 4$ ought to be much smaller ($k_0a < 1$)
than our DNG sphere.

\begin{figure}[tb]
\centering
\epsfig{file=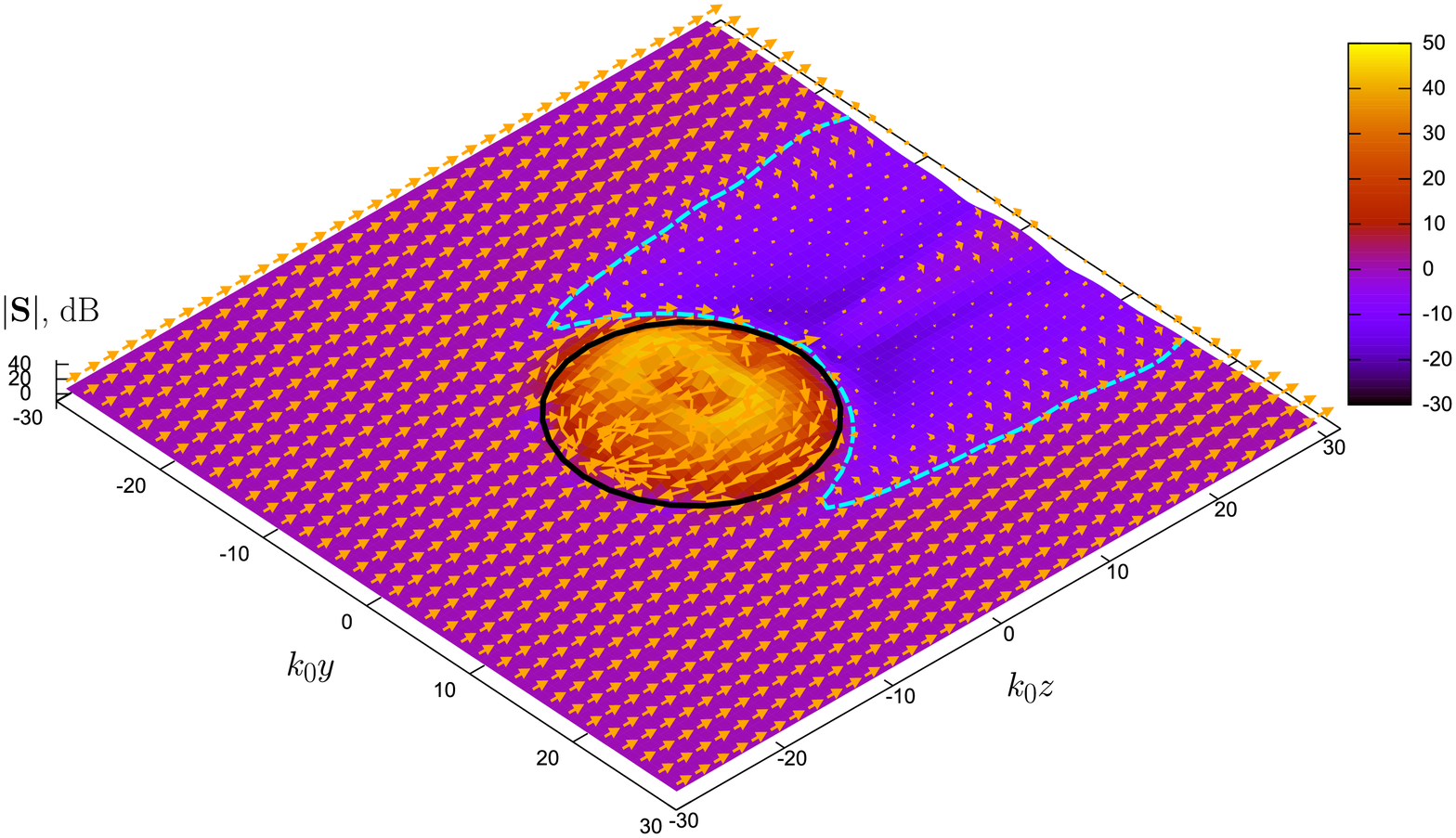,height=0.3\textheight}\\
\epsfig{file=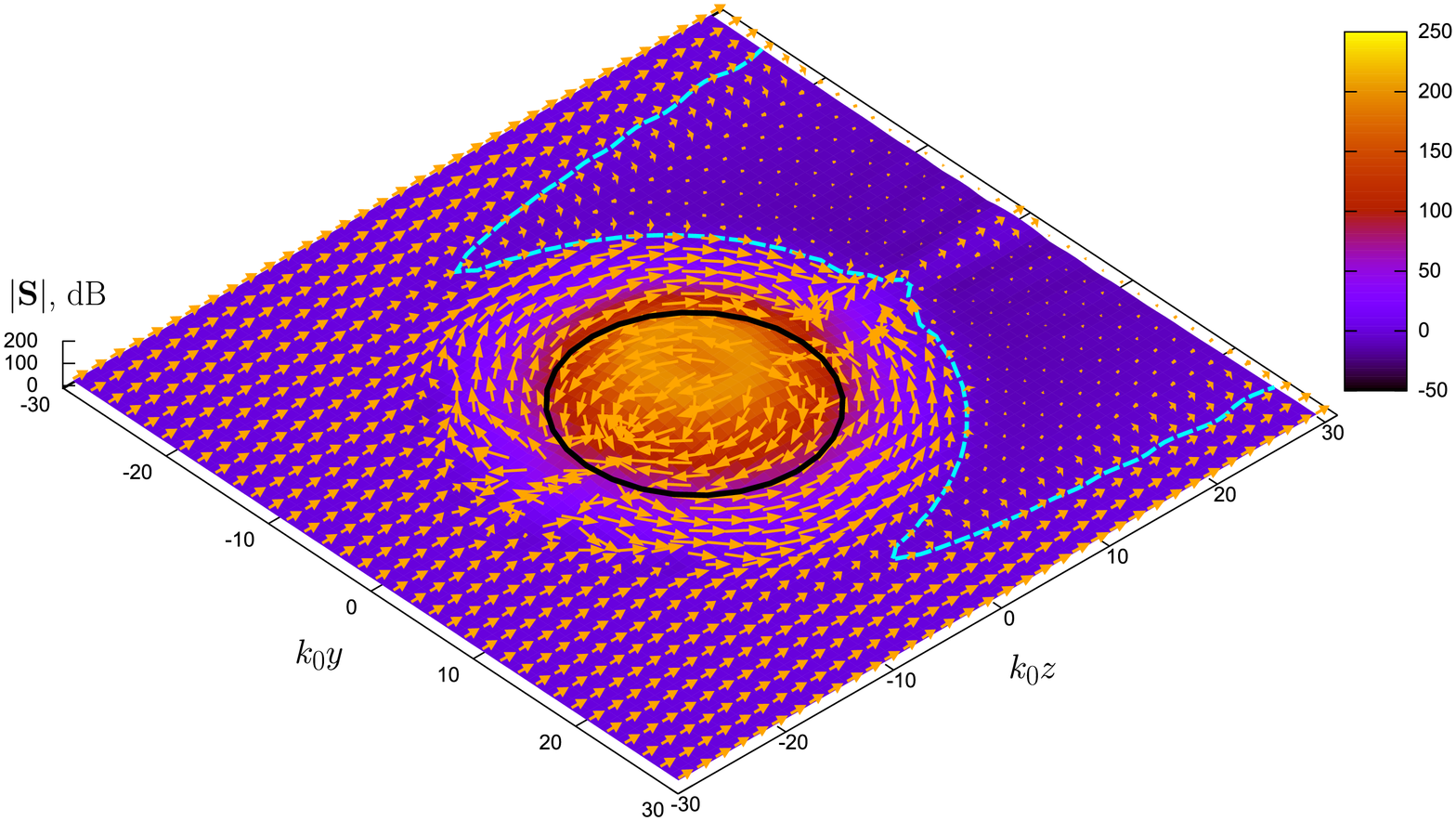,height=0.3\textheight}
\caption{
  \label{Poynting} Top: Magnitude and direction of the Poynting vector $\_S
  =\Re\left(\_E\x{\_H}^*\right)$ inside and outside the metamaterial
  thermal black hole with normalized radius $k_0a = 10$ (the other
  parameters are as in Fig.~\ref{sigma} for $r_2/r_1=10$). The colored
  surface represents the magnitude of the Poynting vector in the
  $yz$-plane relative to the incident plane wave Poynting vector
  magnitude. The arrows indicate the direction of the Poynting
  vector. The thick black circle corresponds to the geometric
  circumference of the object. The dashed cyan line represents the
  shadow region boundary in the half-power criterion. Bottom: the same
  Poynting vector plot as the one shown at the top, but with
  $|\tan\delta| = 10^{-10}$ in the region $r_1<r<r_2$ and whole loss
  concentrated in the core region $r < r_1$, in which case core's
  input impedance is approximately $\eta_0$ for all spherical
  harmonics. In this case, the shadow diameter is about twice as large
  as the diameter of the body.}
\end{figure}

Fig.~\ref{Poynting} shows the power flux density (the Poynting vector)
inside and outside the DNG core-shell object under plane wave
incidence as defined in Sec.~\ref{scattering}.  Note that the power
flux density in regions inside this object is much higher than in the
outside region, which confirms that a metamaterial thermal black hole
is able to greatly concentrate the incident power flux.

The behavior of the Poynting vector at the left and right sides of the
body [i.e., close to the points $(y,z) = (\pm a, 0)$] is especially
peculiar: The flux inside the object is directed oppositely to the
incident flux, as if some power were received from the region of
geometric shadow behind the object.
Similar behavior was observed in
DNG metamaterial waveguides and resonators~\cite{Engheta}.
One can also see from Fig.~\ref{Poynting} that the shadow region has a
larger diameter as compared to the diameter of the body which results
in $\sigma_{\rm abs} > \pi a^2$.  The situation improves even more
when the radius of the core is made smaller. The same figure also
demonstrates the case with a very small value of the loss tangent in
DNG shell region, when all the loss is concentrated within the
core. In this case, a much more pronounced effect is obtained: the
shadow size is about twice the size of the body.

In the second example of Fig.~\ref{sigma} with $r_2/r_1 = 100$, the
normalized absorption cross section attains $\sigma_{\rm abs}/(\pi
a^2) \approx 2.4$ when $k_0a = 10$. Note, however, that in this case
the range of parameter variation is already too large to remain
practical: $\E(r_1)/\E(r_2) = \M(r_1)/\M(r_2) = 10^4$.

For reference, in Fig.~\ref{sigma} we also provide the result for the
case of a uniformly filled sphere, which occurs when $r_1 = r_2$.
For example, for $k_0a = 100$, the uniform DNG sphere with loss tangent
$|\tan\delta| \approx 5\times10^{-2}$ provides $\sigma_{\rm abs}/(\pi
a^2) \approx 1.08$, i.e., even a uniformly filled sphere may
outperform the black body of the same size by about 8\% in this case.

In all examples, the optimum loss tangent value decreases with the
increase of $k_0a$. The values of the normalized absorption cross
section also decrease with $k_0a$. Nevertheless, even when $k_0a =
100$, i.e., when emitter's circumference is 100 wavelength long, we
obtain more then 20\% gain in emitter's performance as compared to
Kirchhoff-Planck's black body of the same size (when $r_2/r_1 = 10$),
and close to 30\% gain when $r_2/r_1 = 100$. The required loss tangent
values in these two cases stay within reasonable limits, for instance,
$|\tan\delta| > 10^{-3}$ for the emitter with $r_2/r_1 = 10$, which
can be realized in an experiment. In our opinion, this is a remarkable
result which shows that even for optically large bodies with $k_0a
\sim 10^2$ there exist emitters which noticeably outperform
Kirchhoff-Planck's black body.

In general, with further decrease in the core radius and $|\tan\delta|$,
the achievable values of $\sigma_{\rm abs}/(\pi a^2)$ become larger,
however, they grow very slowly. Thus, we may conclude that approaching
theoretical result $\sigma_{\rm abs}/(\pi a^2) \rightarrow\infty$ in a
practical DNG emitter will meet with unavoidable obstacles such as
unrealistically high material parameters in the core region $|\E_{\rm
  c}|,\ |\M_{\rm c}|\gg 1$ combined with very low levels of loss:
$|\tan\delta|\ll 1$.

\section{\label{conclusion}Conclusion}

In this paper we have proven, from the point of view of fluctuational
electrodynamics which deals with bodies kept in thermodynamically
equilibrium states and characterized with infinite internal thermal
capacity, that there is no theoretical upper limit on the spectral
power of thermal radiation of finite-size bodies. Thus, the
fluctuational electrodynamics {\em alone} does not set up any bounds
on the level of spectral power emitted by a hot body: for instance, a
conjugate-matched emitter with radius $a\gg \lambda$ can radiate a
much larger (theoretically, infinitely larger) power at the wavelength
$\lambda$ than can be predicted for the same-size body by using
Planck's black body emission formula. This holds at any given
wavelength, and even in a scenario when such a body radiates into
unbounded free space.

This result was obtained by two independent methods: Firstly, by
identifying the conditions which maximize the power emitted by a body
when it performs as a source of thermal radiation, and, secondly, by
maximizing the absorption cross section of a body. As expected, both
derivations lead to the same conclusion stated in the previous
paragraph. Moreover, we have proven that neither the second law of
thermodynamics, nor Kirchhoff's law of thermal radiation (when
properly amended) are violated by theoretical existence of such
strongly super-Planckian emitters.

We have proposed a physical realization of such conjugate-matched
emitters, which employs low-loss media with simultaneously negative
permittivity and permeability --- DNG media. It is known that such
media support strongly resonant surface excitations --- surface
plasmon-polaritons. In flat DNG slabs of infinite extent, these modes
are bound to the surface and the energy associated with them cannot be
emitted into free space. In other words, such modes may not
participate in the far-zone thermal transfer in these geometries (on
the other hand, these modes play the main role in the near-field
super-Planckian thermal transfer). However, the same modes on a curved
closed surface --- like a spherical surface --- always leak some
energy to the free space modes. In our metamaterial superemitter such
leakage is greatly enhanced by tuning the whole structure at resonance
which maximizes the probability of free-space photon emission from
such dark states. Thus, we show that a properly designed metamaterial
structure may dramatically amplify the diffraction effects not
considered in original Planck's theory, and that these effects at a
given frequency can be made even more significant than the standard,
classical effects limited by the geometric optics approximation.

It is quite remarkable that, from a theoretical standpoint, a
finite-size emitter with the double-negative material parameters
derived in Sec.~\ref{realization} performs as a thermal black hole
whose absorption cross section grows without limit when the material
parameters approach the ideal profiles given by Eqs.~\r{DNGeps}
and~\r{DNGmu} with the loss parameter  $\tan\delta\rightarrow 0$. Reciprocally, such
idealized body has infinite effective spectral emissivity when compared to
Planck's black body of a similar size.

However, realizing such a truly super-Planckian emitter in practice
meets with unavoidable obstacles rooted in our inability to obtain DNG
metamaterials with extremely low values of loss and extremely high
absolute values of the permittivity and permeability concentrated in a
tiny region of space. For emitters with practically attainable
parameters, the gain in spectral power is not that high and decreases
when radius-to-wavelength ratio increases. Our numerical examples show
that a practical spherical double-negative core-shell emitter can
outperform Planck's black body (at a given wavelength) by more than
100\% for emitters with circumference on the order of 10 wavelengths,
and by about 20-30\% for emitters with circumference on the order of
100 wavelengths. However, considering wide-spread beliefs that
optically large bodies can never outperform a black body of the same
size when radiating into free space, in our opinion, this still makes
a remarkable achievement even in practical terms.

Finally, let us note that the {\em integral} power emitted {\em at all
  wavelengths} remains sub-Planckian for any body formed by passive
and causal components. This limitation can be readily demonstrated
with the known sum rules for optical scatterers, although this is out
of the scope of this paper.

\ack

S.\ I.\ Maslovski acknowledges financial support under ``Investigador
FCT (2012)'' grant.

\appendix

\section{\label{AppA} Impedances of spherical waves}

The transverse electric and magnetic field components in a spherical
wave harmonic are given by the following expressions:
\[
E_{\theta}^{\rm TE} = {ik_0\over r\sin\theta}{\d U\over \d\varphi}, \quad
E_{\varphi}^{\rm TE} = -{ik_0\over r}{\d U\over \d\theta},
\]
\[
\eta_0H_{\theta}^{\rm TE} = {1\over r}{\d^2 U\over \d r\d \theta}, \quad
\eta_0H_{\varphi}^{\rm TE} = {1\over r\sin\theta}{\d^2 U\over \d r\d
  \varphi},
\]
\[
E_{\theta}^{\rm TM} = {1\over r}{\d^2 V\over \d r\d \theta}, \quad
E_{\varphi}^{\rm TM} = {1\over r\sin\theta}{\d^2 V\over \d r\d
  \varphi},
\]
\[
\eta_0H_{\theta}^{\rm TM} = -{ik_0\over r\sin\theta}{\d V\over \d\varphi}, \quad
\eta_0H_{\varphi}^{\rm TM} = {ik_0\over r}{\d V\over \d\theta},
\]
where $U, V \propto {\cal R}_l(k_0r)Y_l^m(\theta,\varphi)$, for the
outgoing spherical waves propagating towards $r = \infty$, and $U, V
\propto \tilde{\cal R}_l(k_0r)Y_l^m(\theta,\varphi)$, for the incoming
waves propagating towards $r = 0$.

Next, considering, for example, the outgoing TM waves and forming the
ratios of $E_\theta/(\eta_0H_{\varphi})$ and
$E_{\varphi}/(\eta_0H_{\theta})$ we find:
\[
{E_{\theta}^{\rm TM}\over \eta_0H_{\varphi}^{\rm TM}}=
{\d^2 V/(\d r\d\theta)\over ik_0\,{\d V/\d\theta}}=
-i{{\cal
    R}_l'(k_0r)\over {\cal R}_l(k_0r)},
\]
and
\[
{E_{\varphi}^{\rm TM}\over \eta_0H_{\theta}^{\rm TM}}=
-{\d^2 V/(\d r\d\theta)\over ik_0\,{\d V/\d\theta}}=
i{{\cal R}_l'(k_0r)\over {\cal R}_l(k_0r)},
\]
from here we get that the transverse fields in the outgoing waves are
related as $E_\theta^{\rm TM}\hat{\boldsymbol\thetaup} +
E_\varphi^{\rm TM}\hat{\boldsymbol\varphiup}=-Z_{w,nm}^{{\rm
    TM}}\hat{\_r}\x\left(H_\theta^{\rm TM}\hat{\boldsymbol\thetaup} +
  H_\varphi^{\rm TM}\hat{\boldsymbol\varphiup}\right)$, where
$Z_{w,lm}^{{\rm TM}} =-i\eta_0{{\cal R}_l'(k_0r)/{\cal R}_l(k_0r)}$.
Considering in the same manner the outgoing waves of TE polarization,
we obtain $Z_{w,lm}^{{\rm TE}} =i\eta_0{{\cal R}_l(k_0r)/{\cal
    R}_l'(k_0r)}$.

By performing a similar derivation it is straightforward to find that
in the TM-polarized incoming spherical waves the transverse fields are
related as $E_\theta^{\rm TM}\hat{\boldsymbol\thetaup} +
E_\varphi^{\rm TM}\hat{\boldsymbol\varphiup}=\tilde{Z}_{w,nm}^{{\rm
    TM}}\hat{\_r}\x\left(H_\theta^{\rm TM}\hat{\boldsymbol\thetaup} +
  H_\varphi^{\rm TM}\hat{\boldsymbol\varphiup}\right)$, where
$\tilde{Z}_{w,lm}^{{\rm TM}} =i\eta_0{\tilde{\cal
    R}_l'(k_0r)/\tilde{\cal R}_l(k_0r)}$. Respectively, for the
incoming waves of TE polarization we obtain $\tilde{Z}_{w,lm}^{{\rm
    TE}} = -i\eta_0{\tilde{\cal R}_l(k_0r)/\tilde{\cal R}_l'(k_0r)}$.

Note that because $\tilde{\cal R}_l(k_0r) = \left[{\cal
  R}_l(k_0r)\right]^*$, the wave impedances of the incoming and outgoing
waves are related as $\tilde{Z}_{w,lm}^{{\rm TE, TM}} =
\left(Z_{w,lm}^{{\rm TE, TM}}\right)^*$.

Let us now derive the expressions for the reflection coefficients
$\Gamma_{lm}^{\rm TE,TM}$. In the following we suppress the
polarization and modal indices for brevity.

Let us consider a spherical body characterized with the input
admittance $Y_1 = 1/Z_1$, which is under incidence of an incoming
spherical wave with the wave admittance $\tilde{Y}_w =
1/\tilde{Z}_w$. The body reflects this wave with the complex electric
field reflection coefficient $\Gamma$. The reflected outgoing wave has the wave
admittance $Y_w = 1/Z_w$.

In order to find $\Gamma$ we use the boundary condition on body's
surface: $\hat{\_r}\x\_H_{\rm t} = Y_1 \_E_{\rm t}$, where $\_E_{\rm
  t} = E_\theta\hat{\boldsymbol\thetaup} +
E_\varphi\hat{\boldsymbol\varphiup}$ and $\_H_{\rm t} =
H_\theta\hat{\boldsymbol\thetaup} +
H_\varphi\hat{\boldsymbol\varphiup}$ are the total (i.e., incident
plus reflected) transverse electric and magnetic fields: $\_E_{\rm t}
= (1+\Gamma)\_E_{\rm t}^{\rm inc}$, $\hat{\_r}\x\_H_{\rm t} =
(\tilde{Y}_w-\Gamma Y_w)\_E_{\rm t}^{\rm inc}$. From this condition we
obtain
\[
(\tilde{Y}_w-\Gamma Y_w)\_E_{\rm t}^{\rm inc} = Y_1(1+\Gamma)\_E_{\rm
  t}^{\rm inc},
\l{appBC}
\]
Taking into account that $\tilde{Y}_w = {Y_w}^{\!\!\!\!*}\,$, we find
from Eq.~\r{appBC} that $\Gamma = -(Y_1-{Y_w}^{\!\!\!\!*})/(Y_1+Y_w)$ [Eq.~\r{reflTETM}].

The $Z$-parameters for a uniformly filled spherical shell with an
arbitrary inner radius $r_1$ and and arbitrary outer radius $r_2 >
r_1$ can be found in a similar manner by considering partial spherical
waves propagating within the shell in the two opposite directions, and
expressing through these waves the total transverse electric and
magnetic fields at the surfaces $r=r_1$ and $r=r_2$. From these
expressions one obtains the following $Z$-matrix relation between the
electric and magnetic fields at the two sides of the shell:
\[
\vect{
r_1\_E_{\rm t}(r_1)\\
r_2\_E_{\rm t}(r_2)
} =
\matr{
Z_{11} & Z_{12}\\
Z_{21} & Z_{22}
} \.
\vect{
-r_1\hat{\_r}\x\_H_{\rm t}(r_1)\\
r_2\hat{\_r}\x\_H_{\rm t}(r_2)
}
\]
where the four $Z$-parameters: $Z_{11}$, $Z_{12}$, $Z_{21}$, and
$Z_{22}$, are functions of the polar index $l$ only (they do not depend
on the azimuthal index $m$). Because in calculation of the normalized
absorption cross section~\r{sigmaabsDNG} it is enough to consider just
a single polarization, below we provide the final formulas for the
$Z$-parameters for TM polarization:
\begin{eqnarray}
\l{Z11tm}
Z_{11,\,l}^{\rm TM} &=& -i\sqrt{\M\over\E}\,
{\tilde{\cal R}_{l}(kr_2){\cal R}'_{l}(kr_1)-{\cal R}_{l}(kr_2)\tilde{\cal R}'_{l}(kr_1)\over
\tilde{\cal R}_{l}(kr_2){\cal R}_{l}(kr_1)-\tilde{\cal R}_{l}(kr_1){\cal R}_{l}(kr_2)},\\
Z_{12,\,l}^{\rm TM} &=& -i\sqrt{\M\over\E}\,
{\tilde{\cal R}_{l}(kr_1){\cal R}'_{l}(kr_1)-{\cal R}_{l}(kr_1)\tilde{\cal R}'_{l}(kr_1)\over
\tilde{\cal R}_{l}(kr_2){\cal R}_{l}(kr_1)-\tilde{\cal R}_{l}(kr_1){\cal R}_{l}(kr_2)},\\
Z_{21,\,l}^{\rm TM} &=& -i\sqrt{\M\over\E}\,
{\tilde{\cal R}_{l}(kr_2){\cal R}'_{l}(kr_2)-{\cal R}_{l}(kr_2)\tilde{\cal R}'_{l}(kr_2)\over
\tilde{\cal R}_{l}(kr_2){\cal R}_{l}(kr_1)-\tilde{\cal R}_{l}(kr_1){\cal R}_{l}(kr_2)},\\
\l{Z22tm}
Z_{22,\,l}^{\rm TM} &=& -i\sqrt{\M\over\E}\,
{\tilde{\cal R}_{l}(kr_1){\cal R}'_{l}(kr_2)-{\cal R}_{l}(kr_1)\tilde{\cal R}'_{l}(kr_2)\over
\tilde{\cal R}_{l}(kr_2){\cal R}_{l}(kr_1)-\tilde{\cal R}_{l}(kr_1){\cal R}_{l}(kr_2)}.
\end{eqnarray}
In these formulas, $\E$ and $\M$ are the absolute permittivity and the
permeability of shell's material, respectively, $k =
k_0\sqrt{\E\M/(\E_0\M_0)}$. In order to use these formulas in
Eq.~\r{ZinZparam}, one has to set $r_1 = a$ and $r_2 = a^2/a_0$.

\section{\label{AppB}Derivation of $\sigma_{\rm sc}$ and $\sigma_{\rm abs}$}

Due to the presence of a body in the vicinity of the point $r =
0$, the waves~\r{morewaveexp} will scatter on it, producing in this way the scattered field:
\[
\_E^{\rm sc}_{\rm t} =
{1\over k_0r}\sum_{l=1}^\infty\sum_{m=-1,1}\left[C^{\rm TE}_{l,m}\right.
{\cal R}_l(k_0r)\,\_r\x\nabla_{\rm t} +
\left.C^{\rm TM}_{l,m}{\cal
    R}'_l(k_0r)\,r\nabla_{\rm t} \right]Y_l^m(\theta,\varphi).
\l{scatwaveexp}
\]
The amplitudes of the scattered waves $C^{\rm TE,TM}_{l,m}$ can be
found through the complex reflection coefficient for the corresponding
incident spherical harmonic (see~\ref{AppA}):
\[
\Gamma^{\rm TE}_{lm} = -{Y_{1,lm}^{\rm TE}-{Y_{w,lm}^{\rm TE}}^{\!\!\!\!*}\over Y_{1,lm}^{\rm TE}+Y_{w,lm}^{\rm TE}}, \quad
\Gamma^{\rm TM}_{lm} = -{Y_{1,lm}^{\rm TM}-{Y_{w,lm}^{\rm TM}}^{\!\!*}\over Y_{1,lm}^{\rm TM}+Y_{w,lm}^{\rm TM}},
\l{reflTETM}
\]
where we have introduced body's input admittance for a given spherical
harmonic $Y_{1,lm}^{\rm TE,TM} \equiv 1/Z_{1,lm}^{\rm TE,TM}$ and the
wave admittance $Y_{w,lm}^{\rm TE} \equiv 1/Z_{w,lm}^{\rm TE}$. Note a subtle difference
from the more usual reflection formula --- the presence of the complex
conjugate operation, --- which arises from the fact that impedances of
the counter-propagating spherical waves are different and equal the
complex conjugate of each other (see~\ref{AppA}).

Using Eq.~\r{reflTETM},
we may write for TE wave reflections happening at surface $r = a$:
\[
\left(C_{l,m}^{\rm TE} + B_{l,m}^{\rm TE}\right){\cal R}_l(k_0a) =
\Gamma_{lm}^{\rm TE}A_{l,m}^{\rm TE}\tilde{\cal R}_l(k_0a).
\l{rlmte}
\]
Respectively, for the TM waves reflecting at the same surface,
\[
\left(C_{l,m}^{\rm TM} + B_{l,m}^{\rm TM}\right){\cal R}'_l(k_0a) =
\Gamma_{lm}^{\rm TM}A_{l,m}^{\rm TM}\tilde{\cal R}'_l(k_0a).
\l{rlmtm}
\]
Because in the incident plane wave expansion [Eq.~\r{morewaveexp}]
$A_{l,m}^{\rm TE,TM} = B_{l,m}^{\rm TE, TM}$, we find from
Eqs.~\r{rlmte} and~\r{rlmtm} that
\[
C_{l,m}^{\rm TE,TM} = \left(\tilde{\Gamma}_{lm}^{\rm TE,TM} - 1\right)A_{l,m}^{\rm TE,TM},
\]
where
\[
\l{gammatildeTETM}
\tilde{\Gamma}_{lm}^{\rm TE} = {\tilde{\cal R}_l(k_0a)\over{\cal
      R}_l(k_0a)}\Gamma_{lm}^{\rm TE}, \quad
\tilde{\Gamma}_{lm}^{\rm TM} = {\tilde{\cal R}'_l(k_0a)\over{\cal R}'_l(k_0a)}\Gamma_{lm}^{\rm
  TM},
\] are the reflection coefficients with redefined phase such as if
the reflection happened at the point $r = 0$ (note that
$\left|{\tilde{\cal R}_l(k_0a)/{\cal R}_l(k_0a)}\right| =
\left|{\tilde{\cal R}'_l(k_0a)/{\cal R}'_l(k_0a)}\right| = 1$).

The total scattered power is found by integrating the expression for
$\eta_0^{-1}|\_E_{\rm t}^{\rm sc}|^2$ over the closed spherical surface with infinite
radius. In doing so, we use the orthogonality of the Laplace spherical
harmonics, and the fact that
\[
\oint \left|r\nabla_{\rm t} Y_l^m(\theta,\varphi)\right|^2d\Omega =
\oint \left|\_r\x\nabla_{\rm t} Y_l^m(\theta,\varphi)\right|^2d\Omega = l(l+1),
\]
where $d\Omega = \sin\theta\, d\theta\, d\varphi$. We also take into
account that $|{\cal R}_l(k_0r)| = |\tilde{\cal R}_l(k_0r)| \rightarrow 1$
when $r\rightarrow \infty$. Thus, we obtain for the total scattered power
\begin{eqnarray}
\l{scatpower}
P_{\rm sc} &=& {1\over \eta_0k_0^2}\sum_{l=1}^\infty\sum_{m=-1,1}l(l+1)\left(\left|C^{\rm TE}_{l,m}\right|^2 +
\left|C^{\rm TM}_{l,m}\right|^2\right)\nonumber\\
&=&{\pi|E_{\rm inc}|^2\over 4\eta_0k_0^2}\sum_{p=\rm TE, TM}\sum_{l=1}^\infty\sum_{m=-1,1}(2l+1)
\left|1 - \tilde{\Gamma}_{lm}^{\,p}\right|^2.
\end{eqnarray}
From here, the normalized scattering cross section $\sigma_{\rm sc}/(\pi a^2)$
is found as given by Eq.~\r{sigmasc}.

The absorption cross section $\sigma_{\rm abs}$ can be found by
considering the balance of powers delivered to the body by the
incoming waves and taken away by the outgoing waves. The power
associated with each incoming spherical wave is proportional to
$\left|A_{l,m}^{\rm TE,TM}\right|^2$ and the power associated with the
outgoing waves of the same index and polarization is proportional to
$\left|B_{l,m}^{\rm TE,TM}+C_{l,m}^{\rm TE,TM}\right|^2 =
\left|\tilde{\Gamma}_{lm}^{\rm TE,TM}\right|^2\left|A_{l,m}^{\rm
    TE,TM}\right|^2$. The difference of these two amounts represents
the absorbed power. Therefore, the total power absorbed in the body at
a given wavelength can be expressed as [compare with
Eq.~\r{scatpower}]
\begin{eqnarray}
  P_{\rm abs} &=& {1\over
    \eta_0k_0^2}\sum_{p=\rm TE,TM}\sum_{l=1}^\infty\sum_{m=-1,1}l(l+1)\left(\left|A_{l,m}^{p}\right|^2
    -\left|\tilde{\Gamma}_{lm}^{p}\right|^2\left|A_{l,m}^{p}\right|^2\right)\nonumber\\
  &=&{\pi|E_{\rm inc}|^2\over 4\eta_0k_0^2}\sum_{p=\rm TE,
    TM}\sum_{l=1}^\infty\sum_{m=-1,1}(2l+1) \left(1 -
    \left|\tilde{\Gamma}_{lm}^{\,p}\right|^2\right).
\end{eqnarray}
From here we obtain the resulting Eq.~\r{sigmaabs}.

\section{\label{AppC}Coordinate transformation $r\mapsto a^2/r$}

Let us consider Maxwell's equations with isotropic and uniform
material parameters $\E$ and $\M$:
\[
\nabla\x\_E = i\o\mu \_H, \quad
\nabla\x\_H = -i\o\E\_E. \l{appmax}
\]
In spherical coordinates,
\[
\nabla = \hat{\_r}\,{\d\over\d r}+ {1\over
  r}\left(\hat{\boldsymbol{\theta}}{\d\over\d\theta} +
  {\hat{\boldsymbol{\varphi}}\over\sin\theta}{\d\over\d\varphi}\right).
\]
Under a mapping $r = g(r')$, where $g(r')$ is a monotonic function of
$r'$, the nabla operator transforms as
\[
\nabla = {\hat{\_r}\over g'(r')}{\d\over\d r'} + {1\over
  g(r')}\left(\hat{\boldsymbol{\theta}}{\d\over\d\theta} +
  {\hat{\boldsymbol{\varphi}}\over\sin\theta}{\d\over\d\varphi}\right).\l{appnab}
\] The unit vector $\hat{\_r}$ can be expressed as \[ \hat{\_r} = {\nabla
  g(r')\over |\nabla g(r')|} = {g'(r')\over |g'(r')|}{\nabla r'\over
  |\nabla r'|} = {g'(r')\over |g'(r')|}{\hat{\_r}}', \] where $\hat{\_r}'$
is the unit vector in the direction of growth of $r'$. From here and
Eq.~\r{appnab}, \[ \nabla = {1\over |g'(r')|}\left[\hat{\_r}'{\d\over\d
    r'} + \left({r'|g'(r')|\over g(r')}\right){1\over
    r'}\left(\hat{\boldsymbol{\theta}}{\d\over\d\theta} +
    {\hat{\boldsymbol{\varphi}}\over\sin\theta}{\d\over\d\varphi}\right)\right].
\] The expression in the square brackets reduces to the nabla operator
$\nabla'$ in the primed frame $\_r' = (r', \theta, \varphi)$ when the
function $g$ is such that $x\,|g'(x)| = g(x)$, which has two solutions
for monotonic $g(x)$: $g(x) = C x$, and $g(x) = C/x$, where $C$ is an
arbitrary constant.

Only the second possibility is of interest for us. Equating $C = a^2$
we get $r = a^2/r'$ and, respectively, $\nabla = ({r'}^2/a^2)\nabla'$.
Note also that for this transformation $\hat{\_r}' = -\hat{\_r}$, but
$\hat{\boldsymbol{\varphi}}' = \hat{\boldsymbol{\varphi}}$, and
$\hat{\boldsymbol{\theta}}' = \hat{\boldsymbol{\theta}}$. Thus, this
transformation changes a right-handed coordinate system to a
left-handed one. Therefore, the same physical field vectors when
referred from the two coordinate frames are related as $\_E =
(-\hat{\_r}')(-E_{r'}') + \_E_{\rm t}' = \_E'$ and $\_H =
-[(-\hat{\_r}')(-H_{r'}) + \_H_{\rm t}] = -\_H'$. The extra flip in
the sign of the magnetic field is due to the fact that $\_H$ is a
pseudovector (axial vector) which must change sign under an improper
coordinate transformation. Note that this sign change is required in
order to maintain the form of the input impedance expression after the
transformation:
\[
Z_{r>a}^{\rm TE,TM}=-{\_E_{\rm t}\over\hat{\_u}\x\_H_{\rm t}} =
-{\_E'_{\rm t}\over\hat{\_u}'\x\_H'_{\rm t}} = Z_{r'<a}^{\rm TE,TM},
\]
where $\hat{\_u} = \hat{\_r}$ and $\hat{\_u}' = -\hat{\_r}'$ are unit
vectors coincident with the propagation direction of the incident
wave.

By substituting the above relations into Maxwell's
equations~\r{appmax} (note that the curls also flip signs
under an improper transformation) we obtain
\[ \nabla'\x\_E' = i\o\left({\mu a^2\over {r'}^2}\right) \_H', \quad
\nabla'\x\_H' = -i\o\left({\E a^2\over {r'}^2}\right)\_E'. \] The
transformation of the material parameters is apparent from these
equations.

\end{document}